\documentclass[aps]{revtex4}
\usepackage{graphicx}
\usepackage{graphics}
\begin{document}
\title{Instanton Approach to Josephson Tunneling between Trapped Condensates}
\author{Yunbo Zhang$^{1,2}$\footnote[1]{Email:ybzhang@physik.uni-kl.de} and 
H.J.W. M\"uller--Kirsten$^1$\footnote[2]{Email:mueller1@physik.uni-kl.de}}
\affiliation{$^1$Department of Physics, University of Kaiserslautern, D-67653
Kaiserslautern, Germany \\
$^2$Department of Physics and Institute of Theoretical Physics, Shanxi
University, Taiyuan 030006, P.R. China}

\begin{abstract}
An instanton method is proposed to investigate the quantum tunneling between
two weakly--linked Bose--Einstein condensates confined in double--well
potential traps. We point out some intrinsic pathologies in the earlier
treatments of other authors and make an effort to go beyond these very
simple zero order models. The tunneling amplitude may be calculated in the
Thomas-Fermi approximation and beyond it; we find it depends on the number
of the trapped atoms, through the chemical potential. Some suggestions are
given for the observation of the Josephson oscillation and the MQST.

\end{abstract}
\maketitle
\section{Introduction}

The first experimental observations of Bose--Einstein condensation (BEC) in
dilute gases of trapped alkali atoms\cite{Anderson2} have stimulated the
studies of condensates in double well traps. With a fascinating possibility
of the observation of new kinds of macroscopic quantum phenomena\cite
{Dalfovo1}, which are related to the superfluid nature of the condensates,
numerous authors have addressed this subject area both experimentally and
theoretically. Some recent experiments have investigated the relative phase
of two overlapping condensates in different hyperfine states\cite{Hall} and
robust interference fringes between two freely expanding condensates have
been observed\cite{Andrews} after switching off the double--well potential
that confines them, indicating phase coherence both in space and time. In
fact the possibility of condensate tunneling between two adjacent atomic
traps and detection of Josephson-like current phase effects have been
previously suggested\cite{Javanainen1,Dalfovo2,Reinhard} and intensively
studied by two main approaches which are capable of dealing with quantum
tunneling in this variant of the Josephson effect. Authors with a quantum
optics background tend to favor models in which two boson modes are
involved, the so-called two-mode model\cite
{Milburn,Smerzi1,Raghavan1,Raghavan2,Marino,Jack,Javanainen2,Ruostekoski}.
The other category of theories is based on using the differences of
condensate phases and atom numbers between the two sides of the trap as
conjugate quantum variables\cite{Zapata}. A detailed comparison between
these two approaches is given in ref.\cite{Javanainen3}.

The superfluid nature of condensates can be fully tested only through the
observation of superflows. Despite the many experimental efforts being
focused on the creation of a Josephson junction between two condensate
bulks, direct experimental evidences for the atomic oscillation are still
far from being realized. Theoretically the Josephson junction problem has
been studied in the limit of noninteracting atoms for small-amplitude
Josephson oscillations\cite{Javanainen1,Dalfovo2}, including
finite-temperature(damping) effects\cite{Zapata}. Dynamic processes of
splitting a condensate by raising a potential barrier in the center of a
harmonic trap \cite{Javanainen3,Menotti} and decoherence effects and quantum
corrections to the semiclassical mean-field dynamics\cite{Milburn,Marino,
Imamoglu,Vardi,Anglin} have also been studied. It has been pointed
out\cite{Smerzi1} that even though the Bose Josephson Junction (BJJ) is a
neutral-atom system, it can still display the typical ac and dc Josephson
effects occurring in charged Cooper-pair superconducting junctions.
Moreover, a novel nonlinear effect has been predicted to occur in this BJJ:
The self--trapping of a BEC population imbalance arises because of the
interatomic nonlinear interaction in the Bose gas\cite{Milburn,Smerzi1}.
This was considered to be a novel ``macroscopic quantum self--trapping''
(MQST) and was predicted to be observable under certain experimental
conditions. Three related parameters, i.e. the ground state energy $E^0$,
the interaction energy $U$, and more importantly, the tunneling amplitude $K$%
, are still undetermined for a specific geometry of the potential well and
have been taken as constants in refs.\cite{Milburn,Smerzi1}. It is the main
purpose of this paper to present a rigorous derivation of these quantities
and we find that they actually depend on the number of atoms $N$. This $N$%
--dependence refines the conclusions and makes the self--trapping easier to
observe.

We develop the instanton method for a sensitive and precise investigation of
the tunneling between two condensates. The almost trivially looking problem
of the tunneling behavior in a double--well potential has attracted much
attention from theorists for decades. For a single particle, the solution
can be found even in quantum mechanics textbooks\cite{Landau}. The advantage
of a nonperturbative method, as presented here, is that it gives not only a
more accurate description of the tunneling phenomena but also a
comprehensive physical understanding in the context of quantum field theory.
The periodic instanton configurations, which have been shown to be a useful
tool in several areas of research such as spin tunneling\cite{Liang1,Liang2}%
, bubble nucleation\cite{Liang3} and string theory\cite{Park1}, enable also
the investigation of the finite temperature behavior of these systems. In
the case of the Bose--Einstein system, however, we need to evaluate the
tunneling frequency for a finite chemical potential even at zero
temperature, due to the nonlinear interaction between the confined atoms.
Therefore the chemical potential here replaces the position of the excited
energy and gives rise to an expected higher tunneling frequency.

Thus, in this paper we will study the atomic tunneling between two
condensates in a double-well trap, with special attentions paid to the $N-$%
dependent tunneling amplitude. In Sec. II we review the two mode approach
and summarize 3 critical conditions which are related to the MQST effect.
Analytical solutions to the coupled nonlinear BJJ equations are derived in
Sec. III and some novel features about the different modes of MQST are
discussed in detail. In Sec. IV the often used Thomas-Fermi approximation
and the corrections beyond it are used to obtain the two important
parameters in the BJJ tunneling system. By means of the periodic instanton
method, we calculate the tunneling amplitude/frequency between the two
condensates in Sec. V and find they actually depend on the number of the
trapped atoms. Our results can be compared to the simple single-particle
tunneling result in ref.\cite{Milburn} and we find that the latter
corresponds to the low-energy or noninteracting case. A detailed discussion
about the optimal condition for observation of the MQST and the Josephson
tunneling is the subject of Sec. VI where some comments are made to the
maximum amplitude of the oscillation. Finally our summary and conclusions
are given in Sec. VII.

\section{Two-mode Model and Macroscopic Quantum Self-Trapping}

It was realized long ago that the ground state properties of a trapped
Bose-gas can be well described by the Gross-Pitaevskii equation(GPE)\cite
{Gross} 
\begin{equation}
i\hbar \frac{\partial \Phi }{\partial t}=-\frac{\hbar ^2}{2m}\nabla ^2\Phi
+\left[ V_{ext}({\bf r})+g_0\left| \Phi \right| ^2\right] \Phi  \label{gpe}
\end{equation}
where $V_{ext}$ is the external trapping potential and $g_0=\frac{4\pi \hbar
^2a}m$ is the interatomic coupling constant, with $a$, $m$ the atomic
scattering length and mass, respectively. In this paper we will consider
systems interacting with repulsive forces with $a>0$. To obtain the ground
state properties, one can write the macroscopic wave function(or from the
viewpoint of phase transitions, the order parameter) of the condensate as $%
\Phi ({\bf r},t)=\phi ({\bf r})e^{-i\mu t/\hbar },$ where a condensate in
the stationary state was assumed. Then the GPE (\ref{gpe}) becomes 
\begin{equation}
\mu \phi ({\bf r})=\left( H_0+g_0\phi ^2({\bf r})\right) \phi ({\bf r}%
),\quad H_0=-\frac{\hbar ^2\nabla ^2}{2m}+V_{ext}({\bf r})  \label{gpe2}
\end{equation}
with $H_0$ the Hamiltonian for the condensates of noninteracting bosons.
This equation explains the significance of the chemical potential $\mu $ as
an energy of the stationary level and can be used further for calculation of
the coupling between two condensates, say, the tunneling amplitude(transfer
matrix element or Josephson tunneling term) as will be shown below.

Consider a double-well trap produced, for example, by a far-off-resonance
laser barrier that cuts a single trapped condensate into two parts. In the
barrier region the modulus of the order parameter in the GPE is
exponentially small. In other words, the overlap between the condensates
occurs only in the classically forbidden region, where the wave function is
small and nonlinear effects due to interactions can be ignored. Thus we look
for the solution of the time-dependent GPE (\ref{gpe}) with the two-mode
variational ansatz\cite{Smerzi1,Raghavan1} 
\begin{equation}
\Phi ({\bf r},t)=\psi _1(t)\Phi _1({\bf r})+\psi _2(t)\Phi _2({\bf r}).
\label{ansatz}
\end{equation}
Here, the complex coefficients $\psi _i(t)=\sqrt{N_i(t)}\exp [i\theta _i(t)]$
are spatially uniform and contain all of the time dependence, while the two
states $\Phi _1({\bf r})$ and $\Phi _2({\bf r})$ are localized in the left
and right wells, respectively, and contain all of the position dependence.
The total number of atoms is conserved in the macroscopic quantum tunneling
process, i.e., $N_1+N_2=\left| \psi _1\right| ^2+\left| \psi _2\right|
^2=N_T $. In a more accurate theory, one could account for the slow time
evolution of the states $\Phi _i({\bf r})$ due to the mean field interaction
since both the number of particles $N_i(t)$ and phases $\theta _i(t)$ in
each well evolve in time. This, however, can be shown to have a negligible
effect when the amplitude of oscillation is relatively small \cite
{Salasnich1}.

Substituting the two-mode ansatz (\ref{ansatz}) into the GPE (\ref{gpe}),
multiplying by $\Phi _i({\bf r})$ and integrating over position we obtain
the equations of motion for the complex coefficients $\psi _i(t)$, i.e. the
BJJ equations 
\begin{eqnarray}
i\hbar \frac{\partial \psi _1}{\partial t} &=&\left( E_1+U_1N_1\right) \psi
_1-K\psi _2,  \label{bjj} \\
i\hbar \frac{\partial \psi _2}{\partial t} &=&\left( E_2+U_2N_2\right) \psi
_2-K\psi _1,  \nonumber
\end{eqnarray}
where damping and finite temperature effects are ignored. Here $E_{1,2}$ are
the zero-point energies in each well, 
\begin{equation}
E_{1,2}=\int d{\bf r}\Phi _{1,2}({\bf r})H_0\Phi _{1,2}({\bf r}),  \label{e}
\end{equation}
$U_iN_i$ are proportional to the atomic self-interaction energies, with 
\begin{equation}
U_{1,2}=g_0\int d{\bf r}\Phi _{1,2}^4({\bf r})  \label{u}
\end{equation}
and $K$ describes the amplitude of the tunneling between condensates 
\begin{equation}
K=-\int d{\bf r}\Phi _1({\bf r})H_0\Phi _2({\bf r}).  \label{k}
\end{equation}
These quantities, which are expressed in terms of appropriate overlap
integrals of the wave-functions $\Phi _{1,2}({\bf r})$, have been taken as
constants in the previous works and it is one of the main tasks of this
paper to obtain them analytically, provided that a specific geometry of the
trap is given. We further note that the (real) ground state solutions $\Phi
_{1,2}({\bf r})$ for isolated traps with equal population in each well, i.e. 
$N_1=N_2=N_T/2$, satisfy the orthonormality condition 
\begin{equation}
\int d{\bf r}\Phi _i({\bf r})\Phi _j({\bf r})=\delta _{ij}.  \label{orth}
\end{equation}
In the derivation above we neglected terms such as $\int \Phi _1^2\Phi _2^2d%
{\bf r},$ $\int \Phi _1^3\Phi _2d{\bf r,}$ and $\int \Phi _1\Phi _2^3d{\bf r}
$ which have the meanings of higher order overlaps between the condensates.
Attempts to include the mean-field contribution to the coupling between the
condensates have also been made\cite{Williams2}. In refs.\cite{Abdullaev}
the authors studied the coherent atomic oscillations and resonances between
coupled BECs with time-dependent symmetric and asymmetric trapping potential
and oscillating atomic scattering length and an interesting chaotic
macroscopic quantum tunneling phenomenon was found to exist. Our analytical
result for $E$, $U$ and $K$ may be useful in the observation of this
macroscopic quantum effect.

Straightforwardly we can obtain the equations for the relative phase $\phi
(t)=\theta _2(t)-\theta _1(t)$ and fractional population imbalance $z=\left[
N_1(t)-N_2(t)\right] /N_T$ \cite{Smerzi1,Raghavan1} 
\begin{eqnarray}
\dot{z} &=&-\sqrt{1-z^2}\sin \phi ,  \nonumber \\
\dot{\phi} &=&\Lambda z+\frac z{\sqrt{1-z^2}}\cos \phi +\Delta E,
\end{eqnarray}
where we have rescaled $2Kt/\hbar $ to a dimensionless time $t$ for
time-independent coupling $K$ between two condensates. The parameters $%
\Delta E$ and $\Lambda $, which determines the dynamic regimes of the BEC
atomic tunneling, can be expressed as 
\begin{eqnarray}
\Delta E &=&\frac{E_1-E_2}{2K}+\frac{U_1-U_2}{4K}N_T,  \nonumber \\
\Lambda &=&\frac{UN_T}{2K},\qquad U=\left( U_1+U_2\right) /2.  \label{Lambda}
\end{eqnarray}
The canonically conjugated variables $z$ and $\phi $, satisfying the
corresponding Hamilton canonical equations of motion, $\dot{z}=-\frac{%
\partial H}{\partial \phi }$ and $\dot{\phi}=\frac{\partial H}{\partial z}$,
suggest that the total energy of the above system is conserved, 
\[
H=\frac \Lambda 2z^2-\sqrt{1-z^2}\cos \phi +\Delta Ez. 
\]
In a simple mechanical analogy, $H$ describes a nonrigid pendulum of tilt
angle $\phi $ and length proportional to $\sqrt{1-z^2}$, that decreases with
the angular momentum $z$. For simplicity we restrict our calculations to a
symmetric double-well potential, which means that the spatially averaged
quantities (\ref{e}) and (\ref{u}) are equal for each well, allowing us to
make the simplifications $\Delta E=0$. In this case, the equations of motion
reduce to 
\begin{eqnarray}
\dot{z}(t) &=&-\sqrt{1-z(t)^2}\sin \phi (t),  \nonumber \\
\dot{\phi}(t) &=&\Lambda z(t)+\frac{z(t)}{\sqrt{1-z(t)^2}}\cos \phi (t),
\label{zphi}
\end{eqnarray}
with the conserved energy 
\begin{eqnarray}
&&H=H[z(t),\phi (t)]=\frac \Lambda 2z(t)^2-\sqrt{1-z(t)^2}\cos \phi (t) 
\nonumber \\
&=&H[z(0),\phi (0)]=\frac \Lambda 2z(0)^2-\sqrt{1-z(0)^2}\cos \phi (0)=H_0.
\end{eqnarray}

In order to see the oscillations of the fractional population imbalance and
the different kinds of (running- and $\pi $-phase) MQST modes more
transparently, we introduce an effective classical particle whose coordinate
is $z$, moving in a potential $W(z)$ with the initial energy $W_0$. The
effective equation of motion is 
\begin{equation}
\dot{z}(t)^2+W(z)=W_0  \label{wz}
\end{equation}
where 
\begin{eqnarray}
W(z) &=&z^2\left( 1-\Lambda H_0+\frac{\Lambda ^2}4z^2\right) , \\
W_0 &=&1-H_0^2=W[z(0)]+\dot{z}(0)^2.
\end{eqnarray}
Increasing the value of $1-\Lambda H_0$ from negative to positive changes
the effective potential $W(z)$ from a double-well to a parabolic. The motion
in the parabolic potential is Rabi-like oscillation with a zero time-average
value of the fractional population imbalance $z$. For fixed parameters $%
\Lambda $ and $H_0$, the oscillations with small effective energies $W_0$
are sinusoidal, the increasing of the effective energies adds higher
harmonics to the sinusoidal oscillations. In the case of double-well
potential with two minima located at $z_{\pm }=\pm \frac{\sqrt{2(\Lambda
H_0-1)}}\Lambda $ and the barrier height given by $W(0)-W(z_{\pm })=\frac{%
\left( \Lambda H_0-1\right) ^2}{\Lambda ^2}$, the motion is very different
from that in the parabolic potential. If the effective energies is larger
than the barrier between two wells, that is, $W_0>0$ or $H_0<1$, the motion
is a nonlinear Rabi oscillation with a zero time-average value of $z$, which
corresponds to the periodic flux of atoms from one condensate to the other.
If the effective energy is lower than the potential barrier, $W_0<0$ or $%
H_0>1$, the particle is forced to become localized in one of the two wells
and the population in each trap oscillates around a nonzero averaged $%
\left\langle z(t)\right\rangle \neq 0$ which has been termed MQST in \cite
{Smerzi1,Raghavan1}. Based on these analyses, we know the critical condition
for MQST corresponds to the point where the effective energy equals the
potential barrier, $W_0=0$ or $H_0=1$.

The lowest four stationary state solutions ($\dot{z}=\dot{\phi}=0$) of eq.(%
\ref{zphi}) are: {\bf a)} the symmetric ground state $\phi _s=2n\pi $, $%
z_s=0 $ with energy $E_{+}=-1$; {\bf b)} the antisymmetric eigenfunction $%
\phi _s=(2n+1)\pi $, $z_s=0$ with energy $E_{-}=1$; and {\bf c)} the $z-$%
symmetry breaking states $\phi _s=(2n+1)\pi $, $z_s=\pm \sqrt{1-\frac 1{%
\Lambda ^2}}$ with energy $E_{sb}=\frac 12\left( \Lambda +\frac 1\Lambda
\right) >1$, respectively. The degenerate eigenstates that break the $z$
symmetry {\bf c)} are obviously the result of the nonlinear interatomic
interaction. In this sense we may verify again the MQST condition $H_0>1$,
which means the populations become macroscopically self-trapped with $%
\left\langle z\right\rangle \neq 0$. From above the 2 lowest-lying states 
{\bf a)} and {\bf b)} with $z_s=0$ are obviously not MQST($H_0=-1,1$), only
the symmetry breaking states {\bf c)} with $z_s\neq 0$ can be regarded as
being self-trapped ($H_0=\frac 12\left( \Lambda +\frac 1\Lambda \right) >1$).

\begin{figure}[h]
\centering
\includegraphics[totalheight=8cm]{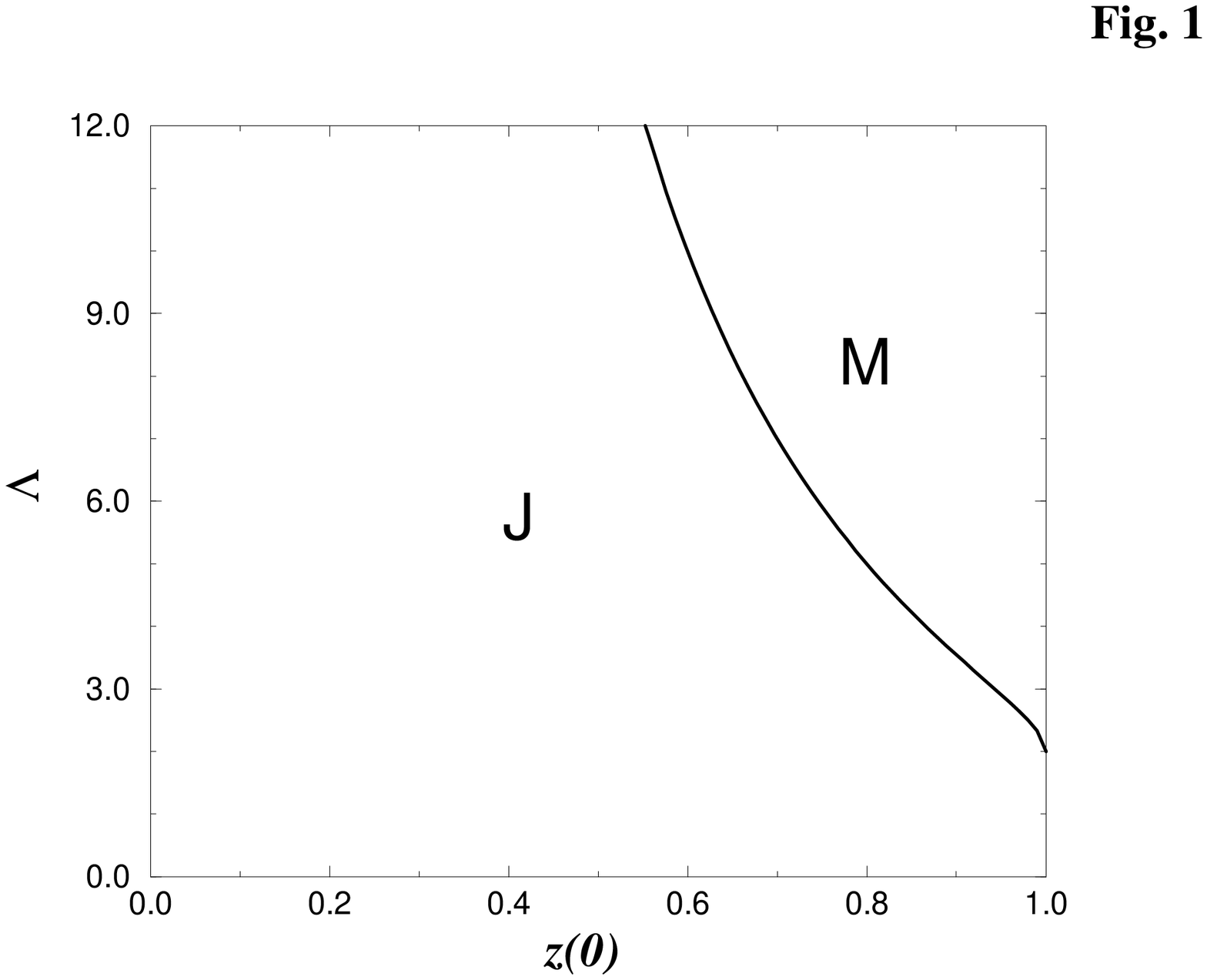}
\caption{The parameter range for the appearance of MQST for $0-$phase
mode: M$\rightarrow $ MQST, J$\rightarrow $ Josephson Tunneling.}
\end{figure}

In summary we observe 3 different critical conditions in refs. \cite
{Smerzi1,Raghavan1}. One is the {\bf MQST condition} which is defined by 
\begin{eqnarray}
&<&z(t)>\neq 0\Rightarrow H_0>1  \nonumber \\
&\Rightarrow &\frac \Lambda 2z^2(0)-\sqrt{1-z^2(0)}\cos \left[ \phi
(0)\right] >1.  \label{mqst}
\end{eqnarray}
In a series of experiments in which $\phi (0)$ and $z(0)$ are kept constant
but $\Lambda $ is varied by changing the geometry or the total number of
condensate atoms, we have the critical parameter 
\begin{equation}
\Lambda _c=\frac{1+\sqrt{1-z(0)^2}\cos \phi (0)}{z(0)^2/2}  \label{lambdac}
\end{equation}
and the critical condition for MQST is $\Lambda >\Lambda _c$. On the other
hand, changing $z(0)$ and keeping $\phi (0)$ and $\Lambda $ constants we
have a critical initial population imbalance $z_c$, which takes the value 
\begin{equation}
z_c=\frac 2\Lambda \sqrt{\Lambda -1}
\end{equation}
in both $0-$phase mode and $\pi -$phase mode case, which describe the
tunneling dynamics with the initial or time-averaged value of the phase
across the junction being $0$ and $\pi $, respectively. But for $\phi (0)=0$%
, if $z(0)>z_c$, MQST sets in, and for $\phi (0)=\pi $, $z(0)<z_c$ marks the
region of MQST. In Fig. 1 we show the parameter range for the appearance of 
MQST and Josephson tunneling for $0-$phase mode, while for the $\pi-$phase mode
this phase diagram manifests much richer structure as shown in Fig. 2. It 
should be pointed out that the vertical axis $\Lambda $ can extend to much 
larger values, but we illustrate in the figure
only the area of interest. For $\Lambda >2$ and $\phi (0)=\pi $, the
effective potential $W(z)$ always has a double-well structure and the system
is self trapped for all values of $z(0)$. We conclude that in the case of
the $\pi -$phase mode the MQST will set in for most of the areas,
leaving only a small area for actual Josephson tunneling.

\begin{figure}[h]
\centering
\includegraphics[totalheight=8cm]{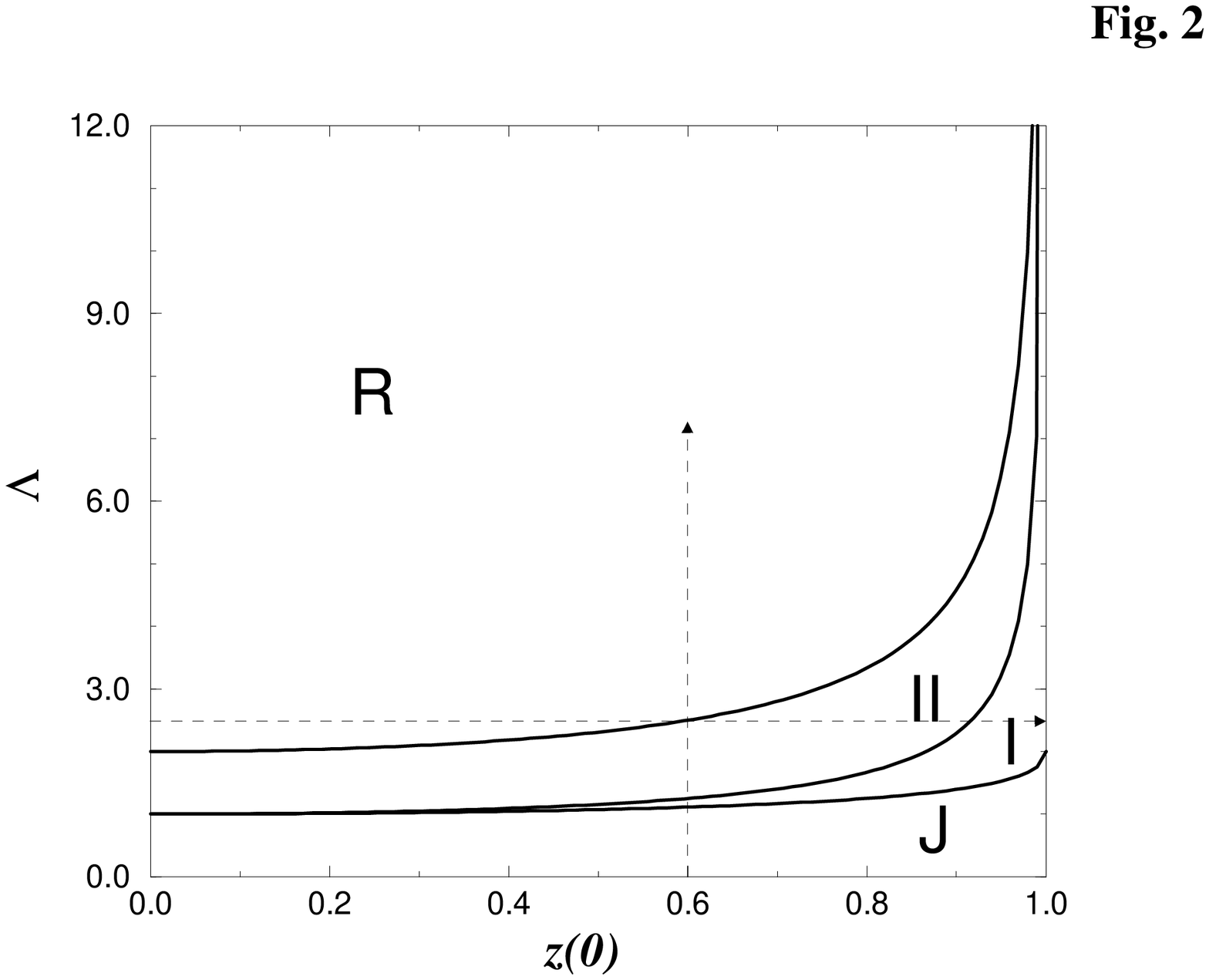}
\caption{Different MQST mode for $\pi-$phase mode: J $\rightarrow $
Josephson Tunneling, I $\rightarrow $ MQST Type I, II $\rightarrow $ MQST
Type II, R $\rightarrow $ MQST unbounded running phase mode. The
vertical and horizontal lines correspond to the case of Fig. 7 and Fig. 5 of
ref. \cite{Raghavan1}, respectively.}
\end{figure}

The {\bf second condition} comes from the stationary $z-$symmetry-breaking
solution {\bf c);} the value of $z_s$ is just the boundary of the first- and
the second-type of trapped states which differ by the time-average value of
the population imbalance $\left\langle z\right\rangle $. We enter into some
detail of the $\pi $ phase mode. According to \cite{Raghavan1}, there are
two kinds of such $\pi $-phase modes with MQST: 
\begin{eqnarray}
\pi \text{-phase modes MQST TypeI} &\Leftrightarrow &\left\langle
z\right\rangle <z_s\neq 0  \label{type12} \\
\pi \text{-phase modes MQST TypeII} &\Leftrightarrow &\left\langle
z\right\rangle >z_s\neq 0  \nonumber
\end{eqnarray}
with $z_s$ being the stationary $z$-symmetry breaking value of the BJJ
equations (\ref{bjj}). Once $\Lambda $ exceeds the value 
\begin{equation}
\Lambda _s=\frac 1{\sqrt{1-z^2(0)}}  \label{lambdas}
\end{equation}
(cf. $z_s=\pm \sqrt{1-\frac 1{\Lambda ^2}}$), a changeover occurs at the
stationary state and the system goes from Type I of the $\pi $-phase trapped
state into Type II. In the phase plane portrait of the dynamical variables $%
z $ and $\phi $ (Fig. 7 in ref.\cite{Raghavan1}) for the fixed value of $%
z(0)=0.6$, this critical condition corresponds to the 4 point-like
trajectories located at ($\pm \pi ,\pm 0.6$). However, in consideration of
the case of a fixed value of $\Lambda $, for example, Figs. 5c and 5d in ref.%
\cite{Raghavan1}, we find that this definition of the Type I and Type II $%
\pi -$phase modes will lead to a dilemma in the analysis of the phase
diagrams and modification to this definition becomes necessary. We return to
this point in the following section.

The {\bf third condition} defines whether the phase $\phi (t)$ is an
unbounded running mode or is localized around $\pi $. Again from the $z\sim
\phi $ phase diagram we know this boundary line can be determined as 
\begin{equation}
\phi |_{z=\pm 1}=\frac \pi 2,\qquad \frac{d\phi }{dz}|_{z=\pm
1}=0\Rightarrow \frac \Lambda 2=H_0
\end{equation}
which gives the critical values 
\begin{equation}
\Lambda _3=\frac 2{\sqrt{1-z^2(0)}},\qquad \text{or}\qquad z_3=\sqrt{1-\frac %
4{\Lambda ^2}}.  \label{lambda3}
\end{equation}
In Fig. 2, we managed to give a phase diagram for the $\pi -$phase mode,
with special attention paid to different MQST modes. Here the area C shows
again the Josephson tunneling area, others are 3 MQST areas, with area I the
bounded MQST of Type I, area II the Type II, and Running Mode for the
unbounded running phase mode, respectively. We used the formulae (\ref
{lambdac}), (\ref{lambdas}) and (\ref{lambda3}) for these three curves, from
the bottom up, respectively. We point out that the second and the third
conditions apply only to the $\phi (0)=\pi $ case.

\section{Exact Analytic Solutions and Different MQST Modes}

From the equation of motion for the effective classical particle (\ref{wz})
we can easily obtain the real time exact solutions in terms of the Jacobian
elliptic functions $%
\mathop{\rm cn}%
$ and $%
\mathop{\rm dn}%
$. This has been done in refs. \cite{Raghavan1,Raghavan2} but there are
some notation errors in the corresponding expressions. We thus rewrite the
solutions here and present a detailed discussion, especially on some
featured oscillation modes. For the parameters in the area of Josephson
tunneling the population imbalance oscillates sinusoidally or
nonsinusoidally between a positive initial value $C$ and a negative one $-C,$
\begin{equation}
z(t)=C%
\mathop{\rm cn}%
\left( \frac{C\Lambda }{2k}(t-t_0),k\right)  \label{cn}
\end{equation}
where 
\begin{eqnarray}
C^2 &=&\frac 2{\Lambda ^2}\left( \left( H_0\Lambda -1\right) +\frac{\zeta ^2}%
2\right) ,  \nonumber \\
\alpha ^2 &=&\frac 2{\Lambda ^2}\left( \frac{\zeta ^2}2-\left( H_0\Lambda
-1\right) \right) ,  \label{amp} \\
\zeta ^2 &=&2\sqrt{\Lambda ^2+1-2H_0\Lambda }  \nonumber
\end{eqnarray}
with $0<k<1$ the elliptic modulus, 
\begin{equation}
k^2=\frac{C^2}{\alpha ^2+C^2}=\frac 12\left( \frac{C\Lambda }\zeta \right)
^2=\frac 12\left( 1+\frac{H_0\Lambda -1}{\sqrt{\Lambda ^2+1-2H_0\Lambda }}%
\right) .
\end{equation}
In the case of MQST, however, the elliptic function $%
\mathop{\rm cn}%
$ will be replaced by its counterpart $%
\mathop{\rm dn}%
$ and the oscillation is about a nonzero average value $\left\langle
z\right\rangle $ 
\begin{equation}
z(t)=C%
\mathop{\rm dn}%
\left( \frac{C\Lambda }2(t-t_0),1/k\right) .  \label{dn}
\end{equation}
From this we know the value $z(t)=0$ is inaccessible at any time and the
modulus is now $1/k$ with $k>1$. The integration constant $t_0$ can be
determined from $z(t)|_{t=0}=z(0):$%
\begin{equation}
t_0=\frac{2k}{C\Lambda }{\rm cn}^{-1}\left( \frac{z(0)}C,k\right) =\frac 2{%
\Lambda \sqrt{\alpha ^2+C^2}}F\left( \arccos \frac{z(0)}C,k\right)
\end{equation}
where $F\left( \phi ,k\right) =\int_0^\phi d\phi (1-k^2\sin \phi )^{-1/2}$
is the incomplete elliptic integral of the first kind. Noting the following
correspondence 
\begin{eqnarray}
0 &<&k<1,\alpha ^2>0,H_0<1,\qquad \text{Josephson Oscillation} \\
k &>&1,\alpha ^2<0,H_0>1,\qquad \qquad \text{MQST}  \nonumber
\end{eqnarray}
we easily derive the physical condition for the onset of MQST, i.e., $H_0=1$
and $\Lambda =\Lambda _c$, from the mathematical condition $k=1$, where the
character of the elliptic function solution changes. The Jacobian elliptic
functions $%
\mathop{\rm cn}%
(u,k)$ and $%
\mathop{\rm dn}%
(u,k)$ are periodic in the argument $u$ with period $4{\cal K}(k)$ and $2%
{\cal K}(k),$ respectively, where ${\cal K}(k)$ is the complete elliptic
integral of the first kind. The time period of the oscillation of $z(t)$ is
then given by 
\begin{eqnarray}
\tau &=&\frac{2k}{C\Lambda }4{\cal K}(k),\quad \text{for}\quad 0<k<1,
\label{tcn} \\
\tau &=&\frac 2{C\Lambda }2{\cal K}(1/k),\quad \text{for}\quad k>1.
\label{tdn}
\end{eqnarray}
In the limit of small amplitude, or linear oscillations, $k\rightarrow 0,$ $%
K(k)\rightarrow \pi /2,$ we have 
\begin{equation}
\tau =\frac{2\pi }{\sqrt[4]{\Lambda ^2+1-2H_0\Lambda }}.
\end{equation}
For $0,\pi -$phase oscillations, $H_0\rightarrow \mp 1,$ the periods in
scaled units $2Kt/\hbar $ become 
\begin{equation}
\tau =\frac{2\pi }{\sqrt{1\pm \Lambda }},
\end{equation}
respectively, which agree with the unscaled ones (eqs (4.7) and (4.10) in
ref.\cite{Raghavan1}) 
\begin{equation}
\tau _{0,\pi }^{-1}=\sqrt{4K^2\pm 2UN_TK}/2\pi \hbar .
\end{equation}

We consider here two special cases, i.e., the $0,\pi -$phase modes. For the $%
0-$phase mode, by inserting $\phi (0)=0$ into eq.(\ref{amp}), we may show
that the oscillation amplitude of $z(t)$ is just the initial value $z(0)$, 
\begin{equation}
C=z(0).
\end{equation}
On the other hand, for the $\pi -$phase mode, this amplitude will depend on
the values of $\Lambda $ and $z(0)$, 
\begin{eqnarray}
C^2 &=&z^2(0),\qquad \qquad \qquad \qquad \qquad \qquad \qquad \text{when }%
\Lambda <\Lambda _s\text{ or }z(0)>z_s,  \label{amppi} \\
C^2 &=&z^2(0)+\frac 4{\Lambda ^2}\left[ \Lambda \sqrt{1-z^2(0)}-1\right]
,\qquad \text{when }\Lambda >\Lambda _s\text{ or }z(0)<z_s.  \nonumber
\end{eqnarray}
In the case of Josephson tunneling for both $0$ and $\pi -$phase modes, that
is, when the oscillation is in the form of elliptic function $%
\mathop{\rm cn}%
$-type, we always have $C=z(0)$. The reason is that under the condition $%
\Lambda >\Lambda _s$ or $z(0)<z_s$ for $\pi -$phase mode the system is
obviously in the region of MQST type II. Therefore we should use the
elliptic function $%
\mathop{\rm dn}%
$ for MQST instead. Generally for MQST the average population imbalance may
be calculated as(the elliptic function $%
\mathop{\rm dn}%
$ oscillates between $1$ and $k_{dn}^{\prime }$) 
\begin{equation}
\left\langle z\right\rangle =\frac C2\left( 1+k_{dn}^{\prime }\right) =\frac %
C2\left( 1+\sqrt{1-1/k^2}\right) =\frac C2\left( 1+\sqrt{1-\frac{2\zeta ^2}{%
C^2\Lambda ^2}}\right) .
\end{equation}
For the $0-$phase mode, $C=z(0)$ and $\zeta ^2=2\left( \Lambda \sqrt{1-z^2(0)%
}+1\right) $, the average population imbalance is 
\begin{equation}
\left\langle z\right\rangle =\frac{z(0)}2\left( 1+\sqrt{1-\frac{4\left(
\Lambda \sqrt{1-z^2(0)}+1\right) }{z(0)^2\Lambda ^2}}\right) .  \label{za0}
\end{equation}
We take the values of Fig. 2 of ref.\cite{Raghavan1} as an example, $\Lambda
=10,z(0)=0.65,\left\langle z\right\rangle =0.46511$, which is exactly the
case. For the $\pi -$phase mode, there are two cases:

{\em Case I:} $\Lambda <\Lambda _s$ or $z(0)>z_s$ we have $C=z(0)$ and $%
\zeta ^2=2\left( 1-\Lambda \sqrt{1-z^2(0)}\right) $which give the average
population imbalance as 
\begin{equation}
\left\langle z\right\rangle =\frac{z(0)}2\left( 1+\sqrt{1-\frac{4\left(
1-\Lambda \sqrt{1-z^2(0)}\right) }{z(0)^2\Lambda ^2}}\right) .  \label{zapi}
\end{equation}
For the parameter in Fig. 3d of ref.\cite{Raghavan1}, $z(0)=0.6,\Lambda
=1.2,\left\langle z\right\rangle =0.54944<0.6$. This is again exactly the
case.

{\em Case II:} $\Lambda >\Lambda _s$ or $z(0)<z_s$ we have 
\begin{eqnarray}
C^2 &=&z^2(0)+\frac 4{\Lambda ^2}\left[ \Lambda \sqrt{1-z^2(0)}-1\right] , \\
\zeta ^2 &=&2\left( \Lambda \sqrt{1-z^2(0)}-1\right) ,  \nonumber
\end{eqnarray}
the average population imbalance is the same as that in Case I eq.(\ref{zapi}%
). For Fig. 3f of ref.\cite{Raghavan1}, $z(0)=0.6,\Lambda =1.3,\left\langle
z\right\rangle =0.63715>0.6$.

We\ may compare our phase diagram Fig. 2 with those previously known
results. First we have a look at the Fig. 7 of ref.\cite{Raghavan1}, which
illustrated the behavior of the system in the $z-\phi $ phase-plane, and
concentrating only on the $\pi -$phase mode. At a fixed value of $z(0)=0.6$,
increasing the value of $\Lambda $ will bring the system from area C to area
I (Type I bounded MQST with $\left\langle z\right\rangle <0.6$), then area
II(Type II bounded MQST with $\left\langle z\right\rangle >0.6$), and
finally into the running mode area. This process is described by a vertical
line in Fig. 2 and it intersects successively the three curves at 
\begin{equation}
\Lambda _c=10/9=1.\dot{1},\qquad \Lambda _s=1.25,\qquad \Lambda _3=2.5,
\end{equation}
from the bottom up. In Table I we list detailed descriptions and graphical
features of the oscillations in accordance with Fig. 7 of ref.\cite
{Raghavan1}. The coincidence here is obvious.

\begin{center}
{\bf Table I: Different oscillation modes for fixed value }$z(0)=0.6${\bf \
\bigskip }

\begin{tabular}{c|c|c}
\hline\hline
$\phi (0)=\pi $ & \multicolumn{1}{||c|}{Descriptions} & \multicolumn{1}{||c}{
Graphics} \\ \hline\hline
$\Lambda =0$ & Rabi oscillation around $\left\langle z\right\rangle =0$ & O
\\ \hline\hline
$0<\Lambda <\Lambda _c$ & sinusoidal/nonsinusoidal $\text{oscillation around 
}\left\langle z\right\rangle =0$ & 0 \\ \hline\hline
$\Lambda =\Lambda _c=1.111$ & the trajectory shrinks and is pinched at the
point $z=0$ & 8 \\ 
& marking the onset of $\pi $-phase MQST type I with $\left\langle
z\right\rangle <z(0)$ &  \\ \hline\hline
$\Lambda _c<\Lambda <\Lambda _s$ & upon further increasing the area are
divided into 2 parts & $_{\circ }^{\circ }$ \\ \hline\hline
$\Lambda =\Lambda _s=1.25$ & the 2 divided areas collapse to 2 point-like
trajectories & {\Huge :} \\ 
& at $z=0.6$; boundary between MQST type I and type II &  \\ \hline\hline
$\Lambda _s<\Lambda <\Lambda _3$ & further increase of $\Lambda $ induces a
reflection of the trajectory & $%
{\text{o} \atop \text{o}}%
$ \\ 
& about the fixed point and $\pi $-phase MQST type II with $\left\langle
z\right\rangle >z(0)$ &  \\ \hline\hline
$\Lambda =\Lambda _3=2.5$ & boundary between bounded-mode(type II) and
running-mode MQST & $%
{\nabla  \atop \Delta }%
$ \\ \hline\hline
$\Lambda >\Lambda _3$ & the trajectory joins the unbounded running-mode MQST
& $%
{\omega  \atop m}%
$ \\ \hline\hline
\end{tabular}
\bigskip
\end{center}

Secondly we go through Fig. 5 of ref.\cite{Raghavan1} to see what happens if
we keep the value of $\Lambda $ constant and increase $z(0)$. This is
indicated in Fig. 2 by a horizontal line. For small values of $z(0),$ the
phase is unbounded and the system exhibits running phase MQST. However,
above a certain value $z_3=\sqrt{1-4/\Lambda ^2}$(cf. $\Lambda _3$; there is
an error in ref. \cite{Raghavan1} where the condition $z(0)=2z_s=2\sqrt{%
1-1/\Lambda ^2}$ is unphysical since it is larger than 1), with $%
\left\langle z\right\rangle $ still nonzero, the phase becomes localized
around $\pi $ and remains bounded for all larger values of $z(0)$. Then the
puzzle appears here. According to ref. \cite{Raghavan1}, in their Figs. 5(c)
and 5(d), $z(0)=0.7$ and $0.98$ mark the two different types of $\pi -$phase
MQST since they are on either side of the stationary state value of $z_s=%
\sqrt{1-1/\Lambda ^2}$. In this case, along the direction of the horizontal
line from left to right, we should enter Type I first ($z(0)=0.7<0.92$),
then Type II($z(0)=0.98>0.92$). But this is not the case, instead we enter
Type II first. The horizontal line cuts the two curves at $z_3=0.6$ and $%
z_s=0.92,$ with the latter corresponding to a point-like trajectory in Fig.
5d of ref. \cite{Raghavan1} at $z=0.92,\phi =\pi .$ Another possibility is
that although the initial value $z(0)=0.7<0.92$ but the average value $%
\left\langle z\right\rangle $ $>0.92$. However, from eq. (\ref{zapi}), we
find that in both cases, i.e. for $z(0)=0.7$ and $0.98$, we have $%
\left\langle z\right\rangle $ $<0.92$.

A solution to this dilemma is to define the MQST types I and II as follows: 
\begin{eqnarray}
\text{MQST Type I } &\Leftrightarrow &\left\langle z\right\rangle <z(0)
\label{type12n} \\
\text{MQST Type II } &\Leftrightarrow &\left\langle z\right\rangle >z(0) 
\nonumber
\end{eqnarray}
instead of (\ref{type12}). Under this novel definition we can explain the
phase diagrams as follows: For $z(0)=0.7$ the average population difference $%
\left\langle z\right\rangle >0.7$ which is obviously Type II, while for $%
z(0)=0.98$ we have $\left\langle z\right\rangle <0.98$ which is type I
according to this definition. Therefore, along the direction of the
horizontal line from left to right, we enter Type II first ($\left\langle
z\right\rangle >z(0)=0.7$), then Type I($\left\langle z\right\rangle
<z(0)=0.98$). At the threshold point, the average value coincides with the
initial value, $\left\langle z\right\rangle =z(0)=z_s=0.92,$ and the
trajectory shrinks to a point. We also notice that the oscillation
trajectories for $z(0)=0.8$ and $0.98$ are the same. However, they belong to
different regions of MQST. In Table II we illustrate the oscillation
behavior of a $\pi -$phase mode when the value of $z(0)$ is increased while
keeping $\Lambda =2.5$, which can be regarded as an additional remark to
Figs. 5c and 5d in ref. \cite{Raghavan1}. A further evidence of the
correctness of the criterion (\ref{type12n}) comes from the fact that from
the analytical result for the average population imbalance in the $\pi -$%
phase mode, that is, eq. (\ref{zapi}), we can derive the second critical
condition for $\Lambda ,$ 
\begin{equation}
\left\langle z\right\rangle =z(0)\Rightarrow \Lambda _s=\frac 1{\sqrt{%
1-z^2(0)}}
\end{equation}
which agrees with eq.(\ref{lambdas}). And in the case of the $0-$phase mode
there is no solution for $\left\langle z\right\rangle =z(0)$ with $%
\left\langle z\right\rangle $ determined by eq.(\ref{za0}), indicating that
the self-trapping can only appear as MQST type I($\left\langle
z\right\rangle <z(0)$).

\begin{center}
\bigskip {\bf Table II: The oscillation behavior of a }$\pi -${\bf phase
mode for fixed value }$\Lambda =2.5\bigskip $

\begin{tabular}{c|c|c|c}
\hline\hline
$z(0)$ & \multicolumn{1}{||c|}{The Potential $W(z)$} & \multicolumn{1}{||c|}{
The Trajectory $\phi $} & \multicolumn{1}{||c}{Descriptions} \\ \hline\hline
$0$ & double-well with $W_0$ & pinched & always \\ 
& just on the top of the barrier & at $z(0)=0$ & MQST \\ \hline\hline
$0<z(0)<z_3$ & $W_0$ under the barrier whose & unbounded & running-phase \\ 
& height increases with $z(0)$ &  & MQST \\ \hline\hline
$z(0)=z_3=0.6$ & barrier height increases & $\phi |_{z=1}=\pi /2,$ & 
boundary between \\ 
& continuously with $z(0)$ & $\frac{d\phi }{dz}|_{z=1}=0$ & unbounded and
localized $\phi $ \\ \hline\hline
$z_3<z(0)<z_s$ & barrier height increases & localized & bounded MQST \\ 
& continuously with $z(0)$ & around $\pi $ & type II with $\left\langle
z\right\rangle >z(0)$ \\ \hline\hline
$z(0)=z_s=0.92$ & barrier attains & shrinks to point-like & boundary between
\\ 
& its maximum height & trajectory at $\pi $ & MQST type I and type II \\ 
\hline\hline
$z_s<z(0)<1$ & barrier height decreases & expands but & bounded MQST \\ 
& with increasing $z(0)$ & again localized & type I with $\left\langle
z\right\rangle <z(0)$ \\ \hline\hline
$z(0)=1$ & barrier height decreases to the & agrees & bounded MQST \\ 
& same height as $z(0)=0.6$ & with $z(0)=0.6$ & type I with $\left\langle
z\right\rangle <z(0)$ \\ \hline\hline
\end{tabular}
\end{center}

\section{Thomas-Fermi Approximation and Beyond}

The macroscopic wave function $\Phi $ associated with the ground state of a
dilute Bose gas confined in the potential $V_{ext}(r)$ obeys the GPE (\ref
{gpe}), which can be obtained alternatively using a variational procedure: 
\begin{equation}
i\hbar \frac \partial {\partial t}\Phi =\frac{\delta E}{\delta \Phi ^{*}},
\end{equation}
where the energy functional $E$ is defined by 
\begin{eqnarray}
E[\Phi ] &=&\int d^3{\bf r}\left[ \frac{\hbar ^2}{2m}\left| \nabla \Phi
\right| ^2+V_{ext}({\bf r})\left| \Phi \right| ^2+\frac{g_0}2\left| \Phi
\right| ^4\right]  \nonumber \\
&=&E_{kin}+E_{ho}+E_{int}.  \label{ef}
\end{eqnarray}
The three terms in the integral are the kinetic energy of the condensate $%
E_{kin}$, the (an)harmonic potential energy $E_{ho}$, and the mean--field
interaction energy $E_{int}$, respectively. By direct integration of the GPE
(\ref{gpe2}) one finds the useful expression 
\begin{equation}
N\mu =E_{kin}+E_{ho}+2E_{int}  \label{mucom}
\end{equation}
for the chemical potential in terms of the different contributions to the
energy functional (\ref{ef}). Further important relationships can be found
by means of the virial theorem 
\begin{equation}
2E_{kin}-2E_{ho}+3E_{int}=0  \label{virial}
\end{equation}
and the thermodynamic definition of the chemical potential 
\begin{equation}
\mu =\partial E/\partial N.  \label{muth}
\end{equation}
It was shown\cite{Dalfovo1,Baym} that the ground state properties of the
trapped Bose gas with repulsive interactions may be described quite
accurately for a sufficiently large number of particles $N$ by a
Thomas-Fermi approximation(TFA), in which the kinetic energy, which is also
usually named quantum pressure, is neglected completely. One then gets the
density profile in the form 
\begin{equation}
n({\bf r})=\phi ^2({\bf r})=g_0^{-1}\left[ \mu -V_{ext}({\bf r})\right]
\label{den}
\end{equation}
in the region where $\mu >V_{ext}({\bf r}),$ and $n({\bf r})=0$ outside. In
the simplest case of an isotropic harmonic trap $V_{ext}(r)=m\omega
_0^2r^2/2 $, the normalization condition on $n({\bf r})$ provides the
relation between chemical potential and number of particles 
\begin{equation}
\mu _{TF}=\frac{\hbar \omega _0}2\left( \frac{15Na}{a_{ho}}\right) ^{2/5}
\label{mutf}
\end{equation}
where the harmonic oscillator length $a_{ho}=(\hbar /m\omega _0)^{1/2}$ is
introduced for simplicity. The density profile (\ref{den}) has the form of
an inverted parabola, which vanishes at the classical turning point ${\bf R}$
defined by the condition $\mu _{TF}=V_{ext}({\bf R})$. For a spherical trap,
this implies $\mu _{TF}=m\omega _0^2R^2/2$ and using result (\ref{mutf}) for 
$\mu _{TF}$, one finds the following expression for the radius of the
condensate 
\begin{equation}
R=a_{ho}\left( \frac{15Na}{a_{ho}}\right) ^{1/5}  \label{rtf}
\end{equation}
which grows with $N$. The energy components assume the following simple
values in the TFA 
\begin{equation}
\frac{E_{kin}}N=0,\qquad \frac{E_{ho}}N=\frac 37\mu _{TF},\qquad \frac{%
E_{int}}N=\frac 27\mu _{TF}.
\end{equation}

The above results are obtained by neglecting the kinetic energy term $%
E_{kin} $ in the GPE and provide an accurate description of the exact
solution in the interior of the atomic cloud where the gradients of the wave
function are small. This approximation does not, however, take properly into
account the decay of the wave function near the outer edge of the cloud, and
the Thomas-Fermi wave function (\ref{den}) consequently leads to unphysical
behavior for some properties, most notably the logarithmically divergence of
the kinetic energy arising at the turning point. A good approximation for
the density in the region close to the classical turning point can be
obtained by a suitable expansion of GPE (\ref{gpe2}\.{)}. In fact, when $%
\left| r-R\right| \ll R$, the trapping potential $V_{ext}(r)$ can be
replaced by a linear ramp, $m\omega _0^2R(r-R)$, and the GPE takes a
universal form\cite{Dalfovo2,Lundh}, yielding the rounding of the surface
profile. Using this procedure it is possible to calculate the kinetic energy
which, in the case of a spherical trap, is found to follow the asymptotic law%
\cite{Dalfovo2} 
\begin{equation}
\frac{E_{kin}}N\simeq \frac 52C,\qquad C=\frac{\hbar ^2}{mR^2}\ln \left( 
\frac R{1.3a_{ho}}\right) .  \label{kin}
\end{equation}
Analogous expansions can be derived for the harmonic potential energy $%
E_{ho} $ and interaction energy $E_{int}$ in the same large $N$ limit\cite
{Fetter}. A straightforward derivation is obtained by using nontrivial
relationships among the various energy components. A first relation is given
by the virial theorem (\ref{virial}). A second one is obtained by using
expressions (\ref{mucom}) and (\ref{muth}) for chemical potential. These two
relationships, together with the asymptotic law (\ref{kin}) for the kinetic
energy, allow one to obtain the expansions 
\begin{equation}
\frac{E_{ho}}N=\frac 37\mu _{TF}+C,\qquad \frac{E_{int}}N=\frac 27\mu
_{TF}-C.
\end{equation}
Correspondingly the chemical potential in TFA is modified beyond TFA as 
\begin{equation}
\mu =\mu _{TF}+\frac 32C.  \label{mub}
\end{equation}

We may now first calculate the two relevant quantities in the Bose Josephson
Junction, i.e., the zero-point energy in each well $E_{1,2}$ in eq.(\ref{e})
and the atomic self-interaction energies $U_{1,2}$ in eq.(\ref{u}) both in
TFA and beyond it, leaving the third one, the tunneling amplitude $K$, to
the next section. We note that in the derivation above, the wave function is
normalized to $N$. If one uses instead a wave function normalized to unity,
the following correspondence should be realized 
\begin{equation}
U_{1,2}N_{1,2}\rightarrow 2\frac{E_{int}}N,\qquad E_{1,2}\rightarrow \frac{%
E_{kin}}N+\frac{E_{ho}}N.
\end{equation}
Setting the initial population difference $z(0)=0$ as in ref.\cite{Smerzi1}%
,i.e., $N_1=N_2=N$, we obtain the ground state energy $E_{1,2}$ and the
interaction self energy $U_{1,2}N_{1,2}$ for the isolated traps beyond the
TFA 
\begin{equation}
E_{1,2}=\frac 37\mu _{TF}+\frac 72C,\qquad U_{1,2}N_{1,2}=\frac 47\mu
_{TF}-2C,  \label{eub}
\end{equation}
while for TFA the correction terms disappear and these energies take simpler
forms 
\begin{equation}
E_{1,2}=\frac 37\mu _{TF},\qquad U_{1,2}N_{1,2}=\frac 47\mu _{TF}.
\label{eutfa}
\end{equation}
Considering a condensate of $N=5000$ sodium atoms confined in a symmetric
spherical trap with frequency $\omega _0=100Hz$, we have the results beyond
the TFA, $E_{1,2}=1.18n{\rm K},U_{1,2}N_{1,2}=1.03n{\rm K}$, and in TFA $%
E_{1,2}=0.9n{\rm K},U_{1,2}N_{1,2}=1.19n{\rm K}$, quite close to the values
estimated in ref.\cite{Smerzi1}.

\section{Calculation of the Tunneling Amplitude between two Trapped
Condensates}

In this section we give a rigorous derivation of the amplitude of the atomic
tunneling at zero temperature between two nonideal, weakly linked
condensates in a double well trap. This induces a coherent, oscillating flux
of atoms between wells, that is a signature of the macroscopic superposition
of states in which the condensates evolve. To date there have been no
reports of experimental observations of Josephson tunneling of a condensate
in a double well trap. Josephson tunneling of a condensate in a one
dimensional optical lattice was reported in ref. \cite{Anderson1} and by
controlling relative strengths of the tunneling rate between traps and
atom-atom interactions within each trap, this technique has been used to
produce atom-number-squeezed states in this lattice potential\cite{Orzel}.
The analogy of the tunneling mechanism in a two-well potential and an array
of wells makes it important to calculate the tunneling amplitude explicitly.

The quantity $K$ in eq.(\ref{k}) corresponds to the tunneling amplitude
which can be calculated by different methods, and we demonstrate in this
work the use of the nonperturbative instanton approach. The nonlinear
interaction between the atoms in the same well will be included, which
modifies only the chemical potential $\mu $ both in and beyond the TFA. It
is easily shown that this tunneling amplitude is just the quantity ${\cal R}$
of \cite{Milburn} (up to a minus sign), if one observes the orthogonality
property of the eigenfunctions eq.(\ref{orth}), with $\Phi _{1,2}({\bf r})$
the local modes in each well. We study the amplitude for tunneling between
the two condensates confined in the wells of an external double--well
potential in the 3-dimensional 
\begin{equation}
V_{ext}({\bf r})=\frac{m\omega _0^2}{8x_0^2}(x^2-x_0^2)^2+\frac 12m\omega
_0^2y^2+\frac 12m\omega _0^2z^2
\end{equation}
where we have assumed as in ref.\cite{Milburn} that the interwell coupling
occurs only along $x$ and the wave function components in the other two
dimensions are orthogonal and contribute a factor of unity. The two minima
are located at $\pm x_0$ on the $x-$axis and the parabolic approximation to
the potential in the vicinity of each minimum is 
\begin{equation}
\tilde{V}^{(2)}({\bf r}-{\bf r}_{1,2})=\frac 12m\omega _0^2\left[ (x\pm
x_0)^2+y^2+z^2\right] .
\end{equation}
The barrier height between the two wells 
\begin{equation}
V_0=\frac 18m\omega _0^2x_0^2  \label{bh}
\end{equation}
is assumed to be high enough so that the overlap between the wave functions
relative to the two traps occurs only in the classically forbidden region
where interaction can be ignored and one can safely use the WKB wave
function approximately \cite{Dalfovo2} 
\begin{equation}
\Phi _{1,2}({\bf r})\sim \frac B{\left\{ 2m\left[ V_{ext}(r)-\mu \right]
\right\} ^{1/4}}\times \exp \left( -\sqrt{\frac{2m}{\hbar ^2}}\int_R^r\left[
V_{ext}(r)-\mu \right] ^{1/2}dr^{\prime }\right)  \label{wkb}
\end{equation}
with $B$ a properly selected coefficient. The direct integration of $K$ in
eq.(\ref{k}) using the above WKB wave functions is quite a difficult task,
as already noticed in ref.\cite{Dalfovo2,Zapata}. In the simplest case the
local mode of each well was assigned in ref. \cite{Milburn} the harmonic
oscillator single particle ground state wave function 
\begin{equation}
\Phi _{1,2}({\bf r})=\left( \frac{m\omega _0}{\hbar \pi }\right) ^{1/4}\exp
\left( -\frac{m\omega _0}{2\hbar }r^2\right)
\end{equation}
and the tunneling frequency was obtained through simple integration of these
Gaussian wave functions 
\begin{equation}
\Omega =\frac 2\hbar {\cal R}=\frac 2\hbar \int d{\bf r}\Phi _1^{*}({\bf r}%
)\left[ V_{ext}({\bf r})-\tilde{V}^{(2)}({\bf r}-{\bf r}_{1,2})\right] \Phi
_2({\bf r})=\frac{x_0^2}{a_{ho}^2}\omega _0e^{-x_0^2/a_{ho}^2}.
\label{fr-mb}
\end{equation}
This result, however, is independent of the characteristic parameters of the
trapped atoms, such as the number of the atoms, the chemical potential and
the interatomic coupling constant, etc. This inadequacy is expected to be
cured by means of the periodic instanton method presented in the following.

Alternatively in some references the authors\cite{Williams2} used the
external double-well potential in the form 
\begin{equation}
V_{ext}({\bf r})=V_H({\bf r})+V_B({\bf r})
\end{equation}
which is created by superimposing a harmonic trap 
\begin{equation}
V_H(\rho ,z)=m\omega _z^2(\lambda ^2\rho ^2+z^2)/2
\end{equation}
and a Gaussian barrier along the $z$ axis 
\begin{equation}
V_B(z)=U_0\exp (-\frac{z^2}{2\sigma ^2}).
\end{equation}
In their calculations, they considered a range of values for the trap
frequencies $\omega _z$ and $\omega _\rho $, the barrier height $U_0$ and
width $\sigma $, and the condensate population $N_c$. In the spherical or
1-dimensional case, one can also consider the time-dependent behavior by
means of the variational ansatz\cite{Menotti}. Our observation is that in
the vicinity of the tunneling region the above potential in the
1-dimensional case 
\begin{equation}
V_{ext}(x)=\frac 12m\omega _0^2x^2+U_0\exp \left( -\frac{x^2}{2\sigma ^2}%
\right)
\end{equation}
can be well approximated by the double-well (\ref{vextx}) considered in this
paper for a variety of the geometric parameters of the trap.

Now we turn to the field theory description of the GPE, i.e. the periodic
instanton method which can not only give the correct exponential
contribution of the Euclidean action but also the prefactor. The equal
population case with $N_1=N_2=N$ is assumed in the following calculations,
due to the negligible effect of the slow time evolution of the number of
particles $N_i(t)$\cite{Salasnich1}. To this end we consider a scalar field
problem in a 1--dimensional time plus 1--dimensional space 
\begin{equation}
V_{ext}(x)=\frac{m\omega _0^2}{8x_0^2}(x^2-x_0^2)^2.  \label{vextx}
\end{equation}
After a Wick's rotation $t=-i\tau $ the Euclidean--Lagrangian equation of
motion for a finite chemical potential takes the form 
\begin{equation}
\frac 12m\left( \frac{dx}{d\tau }\right) ^2-V_{ext}(x)=-\mu .  \label{mo}
\end{equation}
The reason why we can handle a nonlinear problem by means of a linear
equation of motion is that we discuss the tunneling behavior in the barrier
region where the nonlinear interaction is negligibly small. However, there
are obvious differences between the BEC tunneling system and the usual
one-body problem, i.e. the nonlinear interaction contributes a finite
chemical potential, which is just the integration constant on the right hand
side of eq.(\ref{mo}). The classical turning points on both sides of the
barrier can be determined by the relation $V(x_{1,2})=\mu $ as suggested in
ref. \cite{Dalfovo2}. For a noninteracting system the chemical potential
approaches the ground state energy corresponding to the vacuum instanton
case in \cite{Gildener,Liang4}.

Solving this Euclidean time classical equation in the usual way \cite{Liang4}
one obtains the periodic instanton solution in terms of the Jacobian
elliptic function 
\begin{equation}
x_c=2x_0\bar{k}b(\bar{k})/\omega _0%
\mathop{\rm sn}%
\left( b(\bar{k})\tau ,\bar{k}\right)   \label{instanton}
\end{equation}
where $%
\mathop{\rm sn}%
$ is a Jacobian elliptic function with modulus $\bar{k}$, and the parameters
are defined as 
\begin{equation}
b(\bar{k})=\frac{\omega _0}2\sqrt{\frac 2{1+\bar{k}^2}},\qquad \bar{k}^2=%
\frac{1-u}{1+u},\qquad u=\sqrt{\frac \mu {V_0}}  \label{bk}
\end{equation}
with the imaginary time period $T=2{\cal K}(\bar{k})/b(\bar{k})$. The action
for the half period can be calculated along the above instanton trajectory (%
\ref{instanton}) 
\begin{equation}
S_c=\int_{-T/2}^{T/2}d\tau \left( \frac 12m(dx_c/d\tau
)^2+V_{ext}(x_c)\right) =W+\mu T/2
\end{equation}
where 
\begin{equation}
W=\frac 23\frac{8V_0}{\omega _0}(1+u)^{1/2}\left( {\cal E}(\bar{k})-u{\cal K}%
(\bar{k})\right) .  \label{w}
\end{equation}
Here ${\cal E}(\bar{k})$ denotes the complete elliptic integral of the
second kind with modulus $\bar{k}$. The frequency of tunneling between the
two condensates is then given by the energy level splitting of the two
lowest states, i.e. $\Omega =\Delta E/\hbar =2K/\hbar =2{\cal R}/\hbar $ and
can be calculated by means of the path integral method as\cite{Liang4} 
\begin{equation}
\Omega =\frac 1\hbar Ae^{-W/\hbar }=\frac{\sqrt{1+u}}{2K(\bar{k}^{\prime })}%
\omega _0\exp \left[ -\frac W\hbar \right] .  \label{fr-pim}
\end{equation}
We emphasize here that an explicit prefactor $A$ is included in this
formula, which has been proven to be valid for the entire region when the
chemical potential is below the barrier height. A remarkable feature of this
periodic instanton result for the tunneling frequency is that it depends on
the chemical potential $\mu $, or equivalently the number of the trapped
atoms $N$ through eqs. (\ref{mutf}) or (\ref{mub}). This result will affect
the conclusion about the observation of the MQST, as will be shown in next
section. When the chemical potential approaches the top of the barrier, i.e.
\begin{equation}
V_0=\mu 
\end{equation}
the periodic instanton solution (\ref{instanton}) becomes the trivial
configuration $x_c=0$ with a name ``sphaleron'' which means ``ready to fall''%
\cite{Manton}, where a type of quantum-classical transition may occur\cite
{Liang1}. From the eqs.(\ref{mutf}) and (\ref{bh}), in the TFA this means 
\begin{equation}
x_0=2R=2a_{ho}\left( \frac{15N_Ta}{2a_{ho}}\right) ^{1/5}  \label{x01tf}
\end{equation}
where $N_T=N_1+N_2$ is the total number of atoms in both wells together.
Therefore for a specific type of trapped atoms and a given double--well
potential with separation $x_0$ (the number of atoms$N_T$) there exists a critical
number of atoms $N_{c1}$ (critical separation $x_{c1}$) determined by the
above equation, below (above) which the tunneling process will give the main
contribution to the tunneling amplitude. However, above this critical number
of atoms or below this critical separation value another process, i.e. the
over--barrier activation will dominate (which is definitely not ``thermal
activation'' as in spin tunneling since the temperature is zero) (cf. Fig.
3). Between these two processes there exists a crossover. A more explicit
condition for this critical number of atoms (separation between the two
minima) can be derived beyond the TFA from eqs.(\ref{mub}) and (\ref{bh}),
i.e. 
\begin{equation}
x_0=2R\sqrt{1+\frac 35\left( \frac{15N_Ta}{2a_{ho}}\right) ^{-4/5}\ln \left( 
\frac{15N_Ta}{2a_{ho}1.3^5}\right) }.  \label{x01b}
\end{equation}
As an example, we consider two weakly--linked condensates of $N_T=10^4$
sodium atoms, confined in two symmetric spherical traps with frequency $%
\omega _0=100Hz$ as in ref. \cite{Smerzi1}. The critical value for $x_{c1}$
is $x_{c1}=24.58\mu m$ or more accurately $x_{c1}=25.29\mu m$. We note that
here the height of the potential barrier is $V_0=2.21${\rm nK} and the
ground state is located at $\hbar \omega _0/2=0.38${\rm nK} so that there
are several energy levels beneath the barrier height. This means the
interaction between the atoms contributes to the chemical potential, which
effectively raises the classical turning points to a remarkably high level.
Although the atoms remain in the ground state, the interaction energy is so
strong that the vacuum instanton method can no longer be applied. We have to
resort to the periodic instanton method, as will be shown below.

\begin{figure}[h]
\centering
\includegraphics[totalheight=8cm]{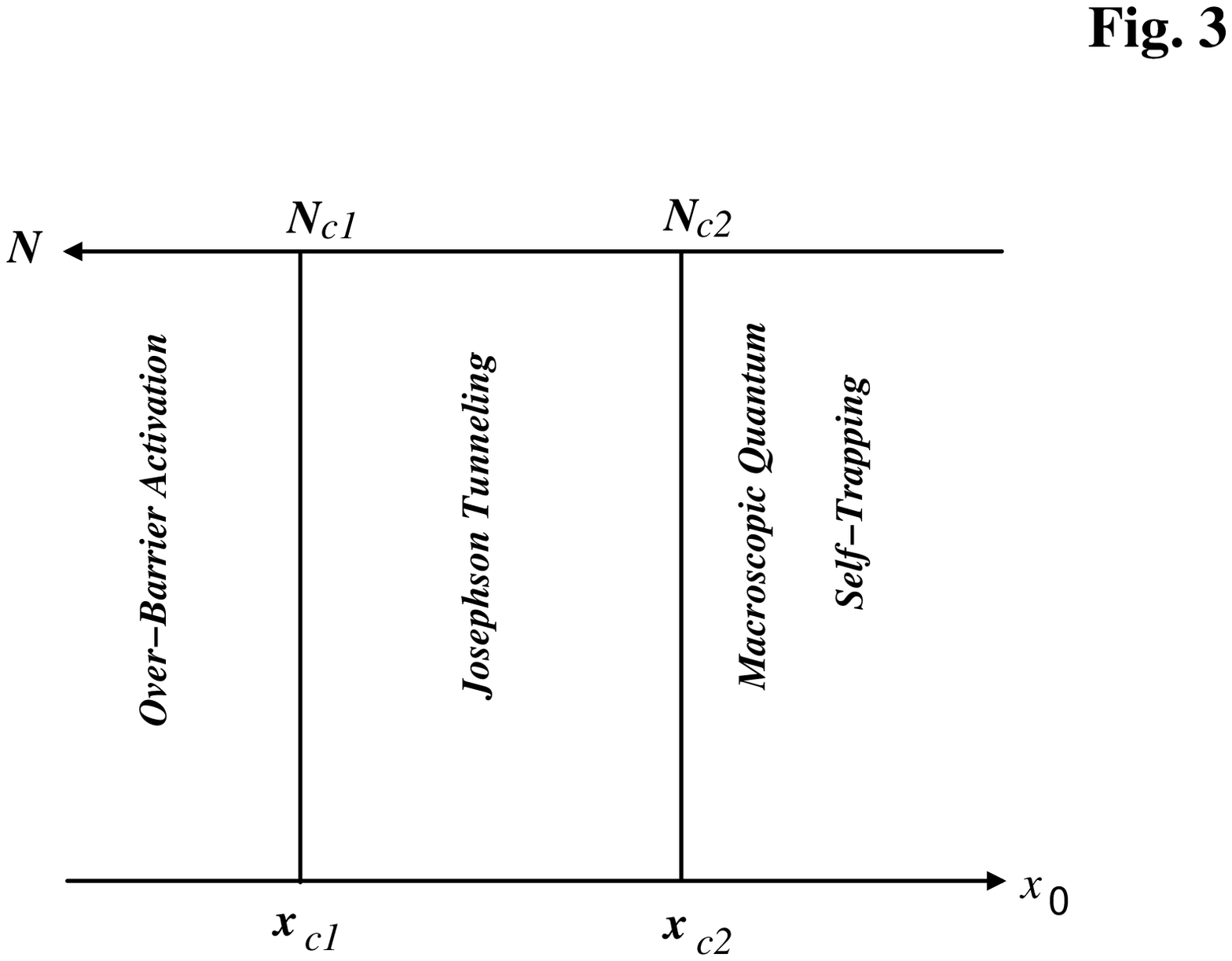}
\caption{3 different processes through which the condensates can
interchange atoms with the corresponding critical parameters of the
boundaries.}
\end{figure}

We now consider the ``low-energy'' or ``non-interacting'' limit, $\mu
\rightarrow 0$. As in the case of a uniform Bose gas, the number of atoms in
the ground state can be macroscopic, i.e., of the order of the total number
in one potential well, when the chemical potential becomes equal to the
energy of the lowest state, which, in our 1--dimensional case here, is $\mu
\rightarrow \mu _c=\frac 12\hbar \omega _0$. The lower boundary for the
chemical potential in fact implies that 
\begin{equation}
\mu _c=\frac 12\hbar \omega _0=\frac 12m_0\omega _0^2R_c^2\rightarrow R_c=%
\sqrt{\frac \hbar {m_0\omega _0}}=a_{ho},
\end{equation}
i.e. the radius of the condensate should never be less than the harmonic
oscillator length $a_{ho}$. We thus have a result similar to that in the
vacuum instanton case\cite{Liang4} and the ``low energy'' limit here is only
meaningful in this sense. Expanding eq. (\ref{fr-pim}) far below the barrier
height, i.e., around the modulus $k\rightarrow 1$, or equivalently
evaluating the tunneling amplitude in the vacuum instanton method \cite
{Liang4}, we obtain for the tunneling frequency 
\begin{equation}
\Omega =2\sqrt{\frac{6S_c}{\pi \hbar }}\omega _0\exp \left( -S_c/\hbar
\right)  \label{fr-vacuum}
\end{equation}
with the Euclidean action 
\begin{equation}
\frac{S_c}\hbar =\frac 23\frac{8V_0}{\hbar \omega _0}=\frac 23\frac{x_0^2}{%
a_{ho}^2}.
\end{equation}
This result can be compared with that of ref. \cite{Milburn}, eq. (\ref
{fr-mb}), which, however, gives not only a smaller exponential contribution 
\begin{equation}
\frac{S_c}\hbar =\frac{8V_0}{\hbar \omega _0}=\frac{x_0^2}{a_{ho}^2},
\end{equation}
(there is a $2/3$ factor) but also an inaccurate prefactor 
\begin{equation}
A=\omega _0S_c/\hbar .
\end{equation}
The source of this inaccuracy is the adoption of the too simple harmonic
oscillator wave function of a single particle, which obviously
oversimplifies the Bose--Einstein condensation tunneling problem. At least
one should use the WKB wave function (\ref{wkb}) in the tunneling region,
and it can be shown that this corresponds to the vacuum instanton result we
present here. For the agreement between WKB and vacuum instanton methods we
refer to ref. \cite{Achuthan}.

We clarify here some features about the two-mode approximation. It is only
when there are many energy levels in each well and the barrier height is
very large that the tunneling behavior can be well defined. As already
pointed out in ref. \cite{Milburn}, the potential should be such that the
two lowest states are closely spaced and well separated from higher levels
of the potential, and that many-particle interactions do not significantly
change this situation. This assumption permits a two-mode approximation to
the many-body description of the system and requires that $x_0\gg a_{ho}$.
In the example considered in ref.\cite{Milburn} the harmonic length is
estimated to be $a_{ho}=1.66\mu m$ for sodium atoms in a double--well trap
with $\omega _0=1kHz$. The two--mode approximation requires that there are
at least two levels beneath the barrier, i.e. 
\begin{equation}
V_0=\frac 18m\omega _0^2x_0^2\gtrsim \frac 32\hbar \omega _0,
\end{equation}
which means the minimum separation $x_0$ should be larger than $5.76\mu m$.
For this separation eq. (\ref{fr-mb}) gives the tunneling frequency $\Omega
\sim 0.0737Hz$, whereas the vacuum instanton result from (\ref{fr-vacuum})
is $\Omega \sim 2.62Hz$; however, it is expected to reach 37\% of $\omega _0$
in ref. \cite{Milburn}, that is $370Hz$(from eq.(\ref{fr-mb}) we know this
frequency can be achieved only at a quite small separation $x_0\sim 1.6\mu m$%
). This is obviously in contradiction. To remedy this, we notice that the
periodic instanton result eq.(\ref{fr-pim}) can reach a maximum frequency $%
\Omega \sim 150Hz$, which is a much higher one but for a quite small number
of trapped atoms $N_{c1}\simeq 535$(Fig. 4). This can be easily understood
as follows. The chemical potential for $5\times 10^5$ atoms in the realistic
experiment of the MIT group in ref. \cite{Anderson2}, $\mu =158.35${\rm nK},
lies much higher than the barrier height $V_0=11.46${\rm nK} if the two
condensates are separated as close as $5.76\mu m$, leading to an unphysical
parameter in eq.(\ref{bk}) $u=\sqrt{\frac \mu {V_0}}$ which is larger than
1. In this case the periodic instanton method cannot apply and the over
barrier activation process will dominate over the Josephson tunneling. In
the language of Josephson junction the two condensates seem to be connected
directly with each other and no barrier exists between them. From another
viewpoint, for the separation smaller than $5.76\mu m$ we cannot put $%
5\times 10^5$ atoms in the double--well, i.e. the barrier seems no longer to
exist for so many atoms and one cannot distinguish to which potential well
the atoms belong and the oscillation is meaningless. The upper boundary for
the trapped number of atoms can be estimated just to be $N_{c1}\simeq 535$
from eq.(\ref{x01b}).

\begin{figure}[t]
\centering
\includegraphics[totalheight=8cm]{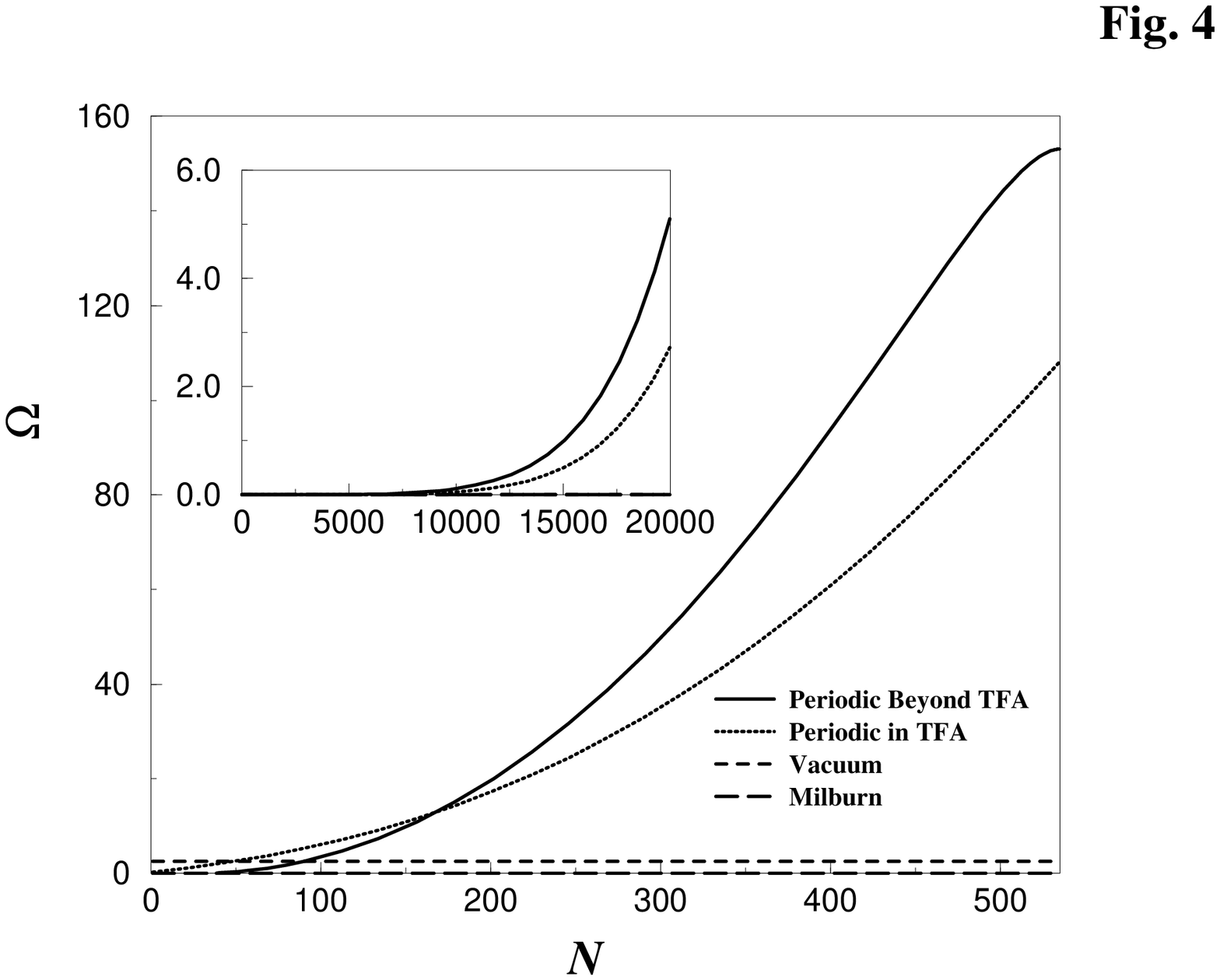}
\caption{ $N-$dependence of the tunneling frequency $\Omega $ for $%
\omega =1kHz,x_0=5.76\mu m$: Solid line is the result of periodic instanton
method beyond the TFA, while dashed line that in TFA. The two horizontal
lines are the results on noninteracting limit, one corresponds to that of
ref.\cite{Milburn}, another is from the vacuum instanton method. Inset: $%
\omega =100Hz,x_0=30\mu m.$ In both cases the constant results are
negligibly small comparing to the $N-$dependent frequencies.}
\end{figure}

In Fig. 4 we show the $N-$dependence of the tunneling frequency $\Omega $
both for a chemical potential in TFA (\ref{mutf}) and beyond it (\ref{mub}).
The periodic instanton results obviously lead to a rapidly growing behavior
for the tunneling frequency when the chemical potential, i.e. the number 
of atoms is increased. The results of ref. \cite{Milburn} eq.(\ref{fr-mb}) and
that from the vacuum instanton method eq.(\ref{fr-vacuum}) are also shown as
horizontal lines.

\section{Observation of Macroscopic Quantum Self Trapping}

We are now in a position to discuss the optimal conditions for observation
of the Josephson oscillation and the MQST effects. According to ref. \cite
{Smerzi1}, for a fixed value of the initial population imbalance $z(0)$ and
phase difference $\phi (0)$, if the parameter $\Lambda $ exceeds a critical
value $\Lambda _c$ in eq.(\ref{lambdac}), the population becomes
macroscopically self--trapped with a nonzero average population difference $%
\left\langle z\right\rangle $. There are different ways in which this state
can be achieved, and all of them correspond to the so-called MQST condition
eq.(\ref{mqst}). Similar results were obtained in ref.\cite{Milburn} for the
case that all atoms are initially localized in one well and it is concluded
that the self-trapping will occur when the total number of the trapped atoms 
$N_T$ exceeds a critical value $N_{c2}$ determined by 
\begin{equation}
N_{c2}=\frac{2\hbar \Omega }{g_0}V_{eff}=\frac{4K}{U_{1,2}}  \label{nc2mb}
\end{equation}
where $V_{eff}^{-1}=\int d{\bf r\Phi }_{1,2}^4({\bf r})$ is the effective
mode volume of each well. For this special case $z(0)=1$, the critical value 
$\Lambda _c$ is shown to be $\Lambda _c=2$ for any initial phase difference $%
\phi (0)$ and eq.(\ref{nc2mb}) is obviously identical with the MQST
condition $UN_T>4K$ from eq.(\ref{mqst}).

The parameters $UN_T$ and $K$, however, are considered as $N-$independent
constants in refs.\cite{Smerzi1,Milburn}. Considering the fact that they
depend actually on the number of the trapped atoms $N$ as in our calculation
above, we can refine the conclusions of refs. \cite{Smerzi1,Milburn}. To
access the region of self--trapping, that is, $\Lambda =\frac{UN_T}{2K}%
>\Lambda _c$, it is better to lower the value of $K$ by making a higher
barrier height $V_0=\frac 18m\omega _0^2x_0^2$ through increasing the
separation $x_0$ or the oscillation frequency $\omega _0$, than to increase
the number of atoms as suggested in ref. \cite{Smerzi1}. In fact, the
quantity $UN_T$ here is proportional to $\mu _{TF}\sim N^{2/5}$ as shown in
eqs.(\ref{eutfa}) and (\ref{mutf}), which means that increasing the number
of atoms will not increase the interaction energy significantly, and at the
same time the tunneling amplitude will be increased more drastically(Fig.
4). Thus, contrary to the observations in ref. \cite{Smerzi1,Milburn}, we
find that the MQST will occur when the number of atoms is smaller (instead
of larger) than a critical value $N_{c2}$, i.e. we should decrease the
number of atoms instead of increasing it (Fig. 3). Inserting the values of
the interaction energy eq.(\ref{eutfa}) or eq.(\ref{eub}), the chemical
potential eq.(\ref{mutf}) or eq.(\ref{mub}) and the tunneling amplitude (\ref
{fr-pim}) into eq.(\ref{mqst}) we may obtain this critical number of atoms
for a given potential geometry in TFA 
\begin{equation}
8\mu _{TF}=7\Lambda _c\hbar \Omega _{TF},
\end{equation}
or beyond it 
\begin{equation}
8\mu _{TF}-28C=7\Lambda _c\hbar \Omega _b
\end{equation}
where $\Omega _{TF}$ and $\Omega _b$ are the tunneling frequencies in eq.(%
\ref{fr-pim}) with the chemical potential $\mu $ taking the values in TFA
and beyond it, respectively.

The parameters which can be adjusted are the number of atoms $N_T$, the
oscillation frequency $\omega _0$, and the separation distance between the
two condensates $x_0$. Fig 3 shows the three different regions for different
numbers of atoms and distances between the potential wells. When $x_0$ ($N_T$%
) is smaller (larger) than the critical value $x_{2c}$($N_{c2}$), the atoms
will oscillate between these two potential wells. Once we increase the
separation above (or decrease the number of atoms below) this critical
value, the MQST will occur, i.e., most of the atoms will tend to remain in
their appropriate wells, leading to only a small oscillation around a fixed
population difference.

\begin{figure}[t]
\centering
\includegraphics[totalheight=8cm]{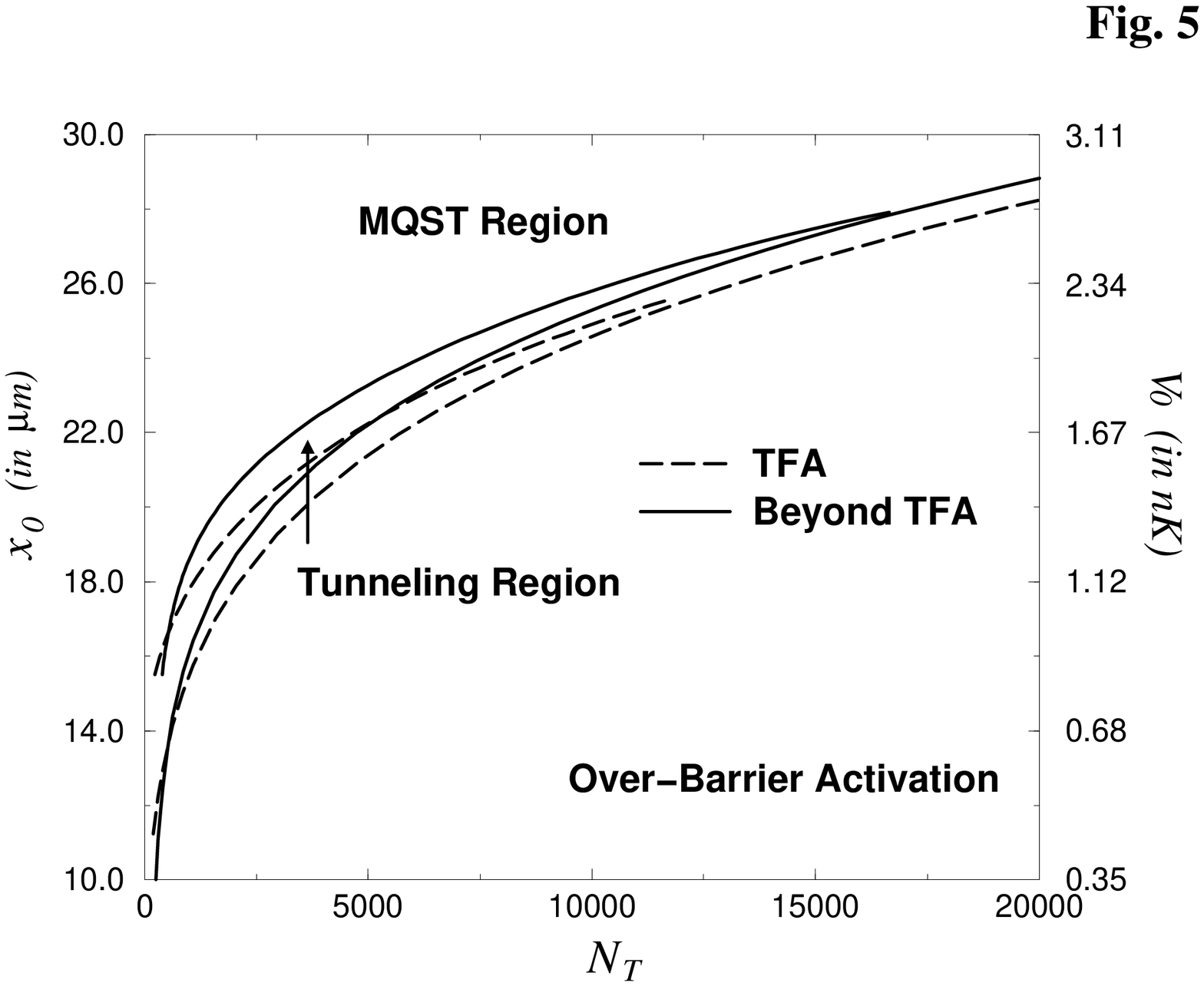}
\caption{ Critical line for MQST effect. Solid line: results beyond TFA
where the parameters take the values $UN_T=8/7\mu _{TF}-4C$ and $\mu =\mu
_{TF}+3C/2$. Dashed line: TFA results where $UN_T=8/7\mu _{TF}$ and $\mu
=\mu _{TF}$ are used in the numerical simulation.}
\end{figure}

We take the initial condition for the population difference to be $z(0)=0.4$
and the zero--phase case $\phi (0)=0$ as an example. Other cases with, for
example, a non--zero phase difference give rise to only a different critical
parameter $\Lambda _c$. For sodium atoms confined in the double--well
potential with $\omega _0=100Hz$, we show numerically in Fig. 5 the critical
line between the three different regions in TFA and beyond it. The upper
region marks the self--trapping region, the lower the over--barrier
activation. Quantum tunneling occurs only for a small range of the
parameter. In the experiment \cite{Andrews}, the barrier was generated by an
off-resonance (blue detuned) laser beam. To make our results more applicable
to experiment we denote on the right vertical axis the corresponding barrier
height in units of {\rm nK}. We also find that the tunneling will be
suppressed when the separation or the number of atoms satisfies $x_0>28\mu m$
or $N_T>12500$ (in TFA $x_0>26\mu m$ or $N_T>12000$). The crossover will
occur directly between the self--trapping and the over--barrier regions,
quite similar to the first--order transition in spin tunneling\cite{Liang1}.
The self trapping region is very easy to access considering the easier
decrease of the tunneling amplitude.

Secondly, from the solution for $z(t)$ in eqs.(\ref{cn}) and (\ref{dn}) we
can define maximum amplitudes of oscillation. For equal populations
initially, i.e. $z(0)=0$, we have $H_0=-\cos \phi (0)$, and the amplitude 
\begin{equation}
C=\frac 1\Lambda \left[ 2\left( \sqrt{\Lambda ^2+1+2\Lambda \cos \phi (0)}%
-\Lambda \cos \phi (0)-1\right) \right] ^{1/2}.  \label{am}
\end{equation}
For $\Lambda >2$, there is only 1 maximum, i.e., $\phi (0)=\pi $. For $%
\Lambda <2,$ however, $\phi (0)=\pi $ is a minimum, while two maxima
correspond to $\cos \phi (0)=-\Lambda /2$. Consequently from eq. (\ref{am})
we have for $\Lambda >2$%
\begin{equation}
z_{\max }=2\sqrt{\Lambda -1}/\Lambda  \label{zmax}
\end{equation}
and for $\Lambda <2$%
\begin{equation}
z_{\max }=1
\end{equation}
We claim here that the result for the maximum amplitude eq.(26) of ref. \cite
{Williams2} is not relevant, in which the author erroneously took $\phi
(0)=\pi /2$ as a maximum. If instead the initial population difference is
nonzero, in ref.\cite{Williams2} the author obtained the same result for the
maximum amplitude as eq.(\ref{zmax}) by setting $\phi (0)=0$ in eq.(\ref
{lambdac}). However, we show here this is not true for $\phi (0)=\pi $
because for the $\pi -$phase mode it is when $z(0)>2\sqrt{\Lambda -1}%
/\Lambda $ that the Josephson oscillation will occur\cite{Raghavan1}. In
this case eq.(\ref{zmax}) gives the minimum amplitude instead of the maximum
one. As already shown in Section III, the oscillation amplitude in the $0-$%
phase mode is just the initial value of population imbalance $z(0)$, while
in the more interesting $\pi -$phase mode, we plot the amplitude eq.(\ref
{amppi}) as a function of $z(0)$ in Fig. 6 for different values of $\Lambda $%
. It is clear that the amplitudes in the $\pi -$phase mode are always larger
than those in the $0-$phase mode. The straight slope line is the result of
the $0-$phase mode. For small values of $\Lambda $, the amplitudes join the
slope line when $z(0)>z_s=\sqrt{1-1/\Lambda ^2}$, with this critical value $%
z_s$ increasing with $\Lambda $, and finally approaching 1 for very large $%
\Lambda $. At a special value of $\Lambda =2,$ it is seen that the
oscillation amplitude can reach 1 even for a zero initial population
imbalance. For very large $\Lambda $ the amplitudes in these two cases agree
tangentially with each other. However, from Fig. 2, the condensates will be
self trapped for any initial population imbalance if $\Lambda >2$, making
the Josephson tunneling unobservable.

\begin{figure}[t]
\centering
\includegraphics[totalheight=8cm]{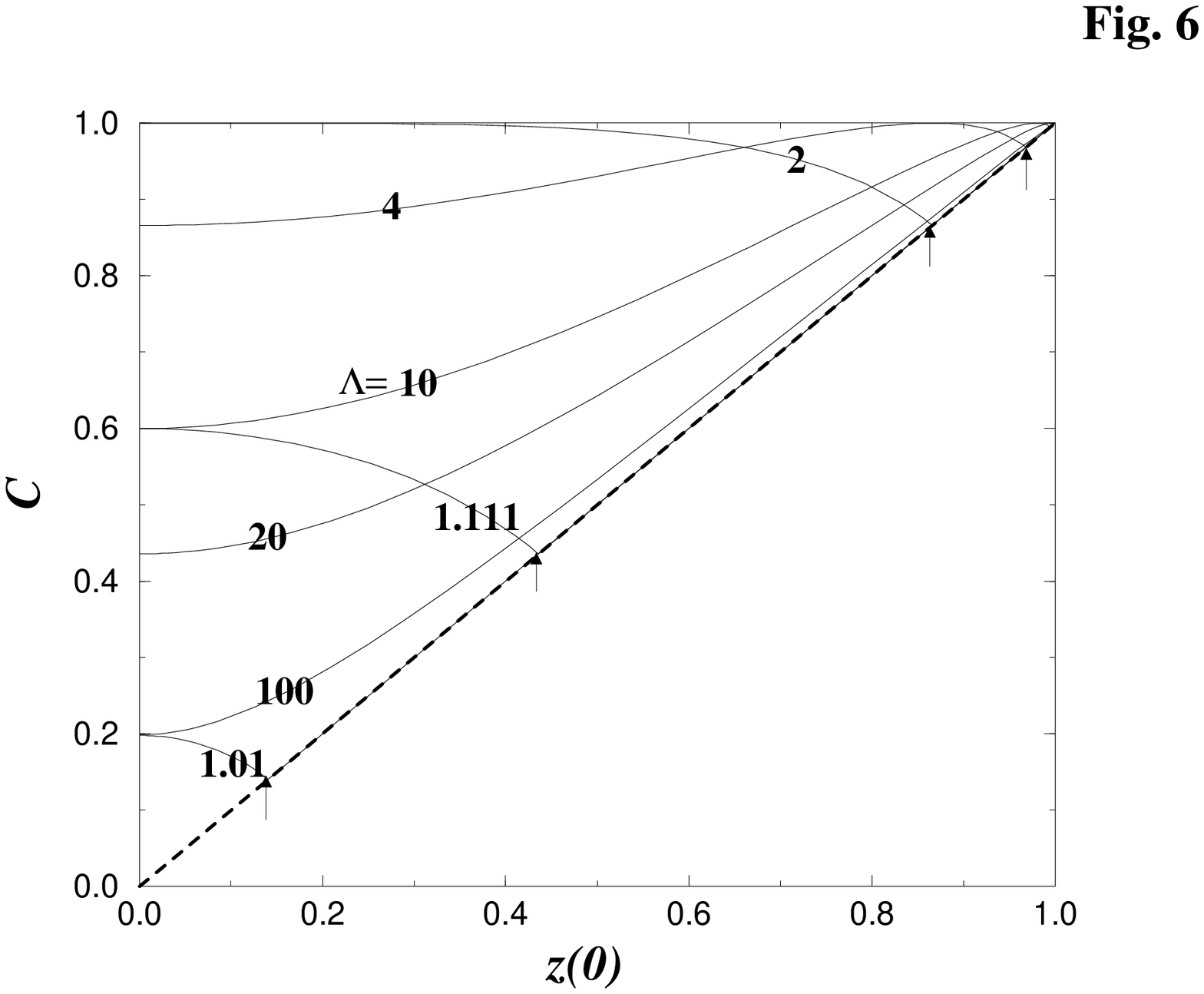}
\caption{ The oscillation amplitude for $0-$ and $\pi -$phase modes.
The slope line is the result of $0-$phase mode, whereas others are results
for different values of $\Lambda $, as indicated in the figure. The arrows
indicate the critical value $z_s$, which increases with $\Lambda $ and
finally approaches 1 for very large $\Lambda $.}
\end{figure}

\section{Conclusions}

We have shown that the periodic instanton method can be used to investigate
the tunneling problem in BEC systems at zero temperature. In particular,
some deficiencies of the earlier treatments are removed. First of all, 
the tunneling amplitude and the nonlinear interaction energy between the 
atoms which have been taken as constants in refs.\cite{Milburn,Smerzi1}, are
calculated analytically in the Thomas--Fermi approximation and beyond it.
The interesting features of the MQST effect are discussed is more detail and
we observe the $N$--dependence of the tunneling amplitude $K$ and the self
interaction energy $UN_T$. Secondly, the $N-$dependence of the tunneling 
frequency $\Omega$ manifests a rapidly growing behavior when the chemical 
potential, i.e., the number of atoms in the trapped condensate is increased.
In this sense, the result of ref. \cite{Milburn} may be considered as that of
a non-interacting approximation. Finnally, for the observations of Josephson 
oscillation and
the self trapping effect, we suggest that one should use a small number of
trapped atoms and change the barrier height through altering the separation $%
x_0$ or the oscillation frequency $\omega _0$ as is evident from the
calculation of this paper. Furthermore the $\pi -$phase mode, that is, when
the relative phase of the wave functions between the two condensates are
opposite to each other, favors the observation of MQST, since a somewhat
large $\Lambda $ will bring the system out of the region of Josephson
tunneling. To observe the Josephson oscillation it is better one adjusts
instead the system into the $0-$phase mode. 

Recently direct observation of an oscillating atomic current has been 
reported in a one-dimensional array of Josephson junctions realized with 
$^{87} Rb$ atomic condensate \cite{Cataliotti}. The authors verified that the 
BEC's dynamics on a lattice is governed by a discrete, nonlinear 
Schr\"{o}dinger equation \cite{Trombettoni} which is common to a large class 
of nonlinear systems. They used a simple variational estimate, assuming a 
gaussian profile for the condensate in each trap, giving the value of 
tunneling amplitude $K$, and a chemical potential $\mu \sim 0.06V_0$ that is much 
lower than the potential barrier $V_0$. This confirms our prediction that 
one should use a small number of trapped atoms for the observations of 
Josephson oscillation. They also observed that the wave functions, as well 
as $K$, depend on the barrier height, however, leaving the analytical result 
unsolved. The instanton result can be extended to the investigation of the 
Bose-gas in such a periodic optical 
lattice \cite{Trombettoni}, the two-component spinor condensates, or even the
metastability in the case of attractive interaction. Work along this
direction is in progress.

\acknowledgments  It is a great pleasure to thank J. E. Williams, F.
Dalfovo, A. Smerzi, X.--B. Wang, W.--D. Li and J.--Q. Liang for useful
discussions. Y.Z. acknowledges support by an Alexander von Humboldt
Foundation Fellowship. This research was supported in part by NSFC of China
under grant No. 10075032.


\begin{references}
\bibitem{Anderson2}  M. H. Anderson, J. R. Ensher, M. R. Matthews, C. E.
Wieman and E. A. Cornell, Science {\bf 269}, 198(1995); K. B. Davis, M.--O.
Mewes, M. R. Andrews, N. J. van Druten, D. S. Durfee, D. M. Kurn and W.
Ketterle, Phys. Rev. Lett. {\bf 75}, 3969(1995); C. C. Bradley, C. A.
Sackett, J. J. Tollett and R. G. Hulet, {\it ibid} {\bf 75}, 1687(1995).

\bibitem{Dalfovo1}  A. S. Parkins, D. F. Walls, Phys. Rep. {\bf 303},
1(1998); F. Dalfovo, S. Giorgini, L. Pitaevskii and S. Stringari, Rev. Mod.
Phys {\bf 71}, 463(1999); A. J. Leggett, Rev. Mod. Phys {\bf 73}, 307(2001).

\bibitem{Hall}  D. S. Hall, M. R. Matthews, C. E. Wieman and E. A. Cornell,
Phys. Rev. Lett. {\bf 81}, 1543(1998); M. R. Matthews, B. P. Anderson, P. C.
Haljan, D. S. Hall, M. J. Holland, J. E. Williams, C. E. Wieman and E. A.
Cornell, {\it ibid} {\bf 83}, 3358(1999).

\bibitem{Andrews}  M. R. Andrews, C. G. Townsend, H.--J. Miesner, D. S.
Durfee, D. M. Kurn and W. Ketterle, Science {\bf 275}, 637(1997).

\bibitem{Javanainen1}  J. Javanainen, Phys. Rev. Lett. {\bf 57}, 3164(1986);
S. Grossmann and M. Holthaus, Z. Naturforsch. Teil A {\bf 50}, 323(1995).

\bibitem{Dalfovo2}  F. Dalfovo, L. Pitaevskii and S. Stringari, Phys. Rev. 
{\bf A54}, 4213(1996).

\bibitem{Reinhard}  W. Reinhard and C. W. Clark, J. Phys. B: At. Mol. Opt.
Phys. {\bf 30}, L785(1997).

\bibitem{Milburn}  G. J. Milburn, J. Corney, E. M. Wright and D. F. Walls,
Phys. Rev. {\bf A55}, 4318(1997).

\bibitem{Smerzi1}  A. Smerzi, S. Fantoni, S. Giovanazzi and S. R. Shenoy,
Phys. Rev. Lett. {\bf 79}, 4950(1997).

\bibitem{Raghavan1}  S. Raghavan, A. Smerzi, S. Fantoni and S. R. Shenoy,
Phys. Rev. {\bf A59}, 620(1999).

\bibitem{Raghavan2}  S. Raghavan, A. Smerzi, S. Fantoni and S. R. Shenoy,
Int. J. Mod. Phys. {\bf B13}, 633(1999).

\bibitem{Marino}  I. Marino, S. Raghavan, S. Fantoni, S. R. Shenoy and A.
Smerzi, Phys. Rev. {\bf A60}, 487(1999); S. Raghavan, A. Smerzi and V. M.
Kenkre, {\it ibid }{\bf 60}, R1787(1999); A. Smerzi and S. Raghavan, {\it %
ibid} {\bf 61}, 063601(2000); S. Giovanazzi, A. Smerzi and S. Fantoni, Phys.
Rev. Lett. {\bf 84}, 4521(2000).

\bibitem{Jack}  M.W. Jack, M.J. Collett and D.F. Walls, Phys. Rev. {\bf A54}%
, R4625(1996).

\bibitem{Javanainen2}  J. Javanainen and M. Wilkens, Phys. Rev. Lett. {\bf 78%
}, 4675(1997); A.J. Leggett and F. Sols, {\it ibid} {\bf 81}, 1344(1998); J.
Javanainen and M. Wilkens, {\it ibid} {\bf 81}, 1345(1998).

\bibitem{Ruostekoski}  J. Ruostekoski and D. F. Walls, Phys. Rev. {\bf A58},
R50(1998); K. Molmer, {\it ibid} {\bf 58}, 566(1998); E. A. Ostrovskaya, Y.
Kivshar, M. Lisak, B. Hall, F. Cattani and D. Anderson, {\it ibid} {\bf 61},
031601(2000).

\bibitem{Zapata}  I. Zapata, F. Sols and A. J. Leggett, Phys. Rev. {\bf A57}%
, R28(1998); A .J. Leggett and F. Sols, Found. Phys. {\bf 21}, 353(1991); F.
Sols, Physica {\bf B 194-196}, 1389(1994); S. Kohler and F. Sols, Phys. Rev. 
{\bf A63}, 053605(2001).

\bibitem{Javanainen3}  J. Javanainen and M. Yu. Ivanov, Phys. Rev. {\bf A60}%
, 2351(1999).

\bibitem{Menotti}  C. Menotti, J. R. Anglin, J. I. Cirac and P. Zoller,
Phys. Rev. {\bf A63}, 023601(2001); R. W. Spekkens and J. E. Sipe, {\it ibid}
{\bf 59}, 3868(1999);

\bibitem{Imamoglu}  A. Imamoglu, M. Lewenstein and L. You, Phys. Rev. Lett. 
{\bf 78}, 2511(1997); Y. Castin and J. Dalibard, Phys. Rev. {\bf A55},
4330(1997); R. Graham, T. Wong, M. J. Collett, S. M. Tan and D. F. Walls, 
{\it ibid} {\bf 57}, 493(1998); J. I. Cirac, M. Lewenstein, K. Molmer and P.
Zoller, {\it ibid} {\bf 57}, 1208(1998); M. J. Steel and M. J. Collett, {\it %
ibid} {\bf 57}, 2920(1998); P. Villain and M. Lewenstein, {\it ibid} {\bf 59}%
, 2250(1999); F. Meier and W. Zwerger, cond--mat/9904147.

\bibitem{Vardi}  A. Vardi and J. R. Anglin, Phys. Rev. Lett. {\bf 86},
568(2001); L. Pitaevskii and S. Stringari, cond--mat/0104458.

\bibitem{Anglin}  J. R. Anglin, P. Drummond and A. Smerzi, cond--mat/0011440.

\bibitem{Landau}  L. D. Landau and E. M. Lifshitz, {\it Quantum Mechanics,
3rd ed.} (Pergamon Press, Oxford, 1977); E. Merzbacher, {\it Quantum
Mechanics, 3rd ed.}(John Wiley Sons Inc, 1998).

\bibitem{Liang1}  J.--Q. Liang, H. J. W. M\"{u}ller--Kirsten, D. K. Park and
F. Zimmerschied, Phys. Rev. Lett. {\bf 81}, 216(1998).

\bibitem{Liang2}  J.--Q. Liang, Y.--B. Zhang, H. J. W. M\"{u}ller--Kirsten,
J. G. Zhou, F. Zimmerschied and F.--C. Pu, Phys. Rev. {\bf B57}, 529(1998).

\bibitem{Liang3}  J.--Q. Liang, H. J. W. M\"{u}ller--Kirsten, Y.--B. Zhang,
A. V. Shurgaia, S. P. Kou and D. K. Park, Phys. Rev. {\bf D62}, 025017(2000).

\bibitem{Park1}  D. K. Park, H. J. W. M\"{u}ller--Kirsten and J.--Q. Liang,
Nucl. Phys. {\bf B578}, 728(2000).

\bibitem{Gross}  L. Pitaevskii, Sov. Phys. JETP {\bf 13}, 451(1961); E. P.
Gross, Nuovo Cimento {\bf 20}, 454(1961); J. Math. Phys. {\bf 4}, 195(1963).

\bibitem{Salasnich1}  L. Salasnich, A. Parola and L. Reatto, Phys. Rev. {\bf %
A60}, 4171(1999); J. Williams, R. Wasler, I. Cooper, E. Cornell and M.
Holland, {\it ibid} {\bf 59}, R31(1999)..

\bibitem{Williams2}  J. Williams, Phys. Rev. {\bf A64}, 013610(2001)

\bibitem{Abdullaev}  F. K. Abdullaev and R. A. Kraenkel, Phys. Rev. {\bf A62}%
, 023613(2000); cond--mat/0004117 and cond--mat/0005445

\bibitem{Baym}  G. Baym and C. J. Pethick, Phys. Rev. Lett. {\bf 76},
6(1996).

\bibitem{Fetter}  A. L. Fetter and D. L. Feder, Phys. Rev. {\bf A58},
3185(1998).

\bibitem{Lundh}  E. Lundh, C. J. Pethick and H. Smith, Phys. Rev {\bf A55},
2126(1997).

\bibitem{Anderson1}  B. P. Anderson and M. A. Kasevich, Science {\bf 282},
1686(1998).

\bibitem{Orzel}  C. Orzel, A. K. Tuchman, M. L. Fenselau, M. Yasuda and M.
A. Kasevich, Science {\bf 291}, 2386(2001).

\bibitem{Gildener}  E. Gildener and A. Patrascioiu, Phys. Rev. {\bf D16},
423(1977).

\bibitem{Liang4}  J.--Q. Liang and H. J. W. M\"{u}ller--Kirsten, Phys. Rev. 
{\bf D46}, 4685(1992).

\bibitem{Manton}  N. S. Manton and T. S. Samols, Phys. Lett. {\bf B207},
179(1988).

\bibitem{Achuthan}  P. Achuthan, H. J. W. M\"{u}ller--Kirsten and A.
Wiedemann, Fortschr. Phys. {\bf 38}, 77(1990).

\bibitem{Cataliotti}  F. S. Cataliotti, S. Burger, C. Fort, P. Maddaloni, 
F. Minardi, A. Trombettoni, A. Smerzi, M. Inguscio, Science {\bf 293}, 884(2001).

\bibitem{Trombettoni}  A.Trombettoni and A. Smerzi, Phys. Rev. Lett. {\bf 86}, 
2353(2001).

\end{references}
\end{document}